\documentclass[%
reprint,
aps,
pra,
letterpaper,
twocolumn,
]{revtex4-2}

\usepackage[utf8]{inputenc}
\usepackage{amsmath,amsfonts,amssymb}
\usepackage{mathtools}
\usepackage{bm}
\usepackage[caption=false,justification=justified]{subfig}
\usepackage[colorlinks,unicode]{hyperref}
\usepackage{enumitem}
\usepackage{dsfont}
\usepackage[dvipsnames]{xcolor}
\usepackage{tikz}
\usetikzlibrary{shapes}
\usetikzlibrary{quantikz}

\newcommand{\eye}{\operatorname{i}}
\newcommand{\op}[1]{\ensuremath{#1}}
\newcommand{\RomanNumeralCaps}[1]{\MakeUppercase{\romannumeral #1}}
\newcommand{\map}[1]{\ensuremath{\mathbf{#1}}}
\newcommand{\textgate}[1]{\text{\textsc{#1}}}
\newcommand{\vac}[1]{\ensuremath{\tilde{#1}}}
\newcommand{\e}{\operatorname{e}}
\newcommand{\aket}[1]{\big| #1 \big>} 
\newcommand{\abra}[1]{\big< #1 \big|} 
\newcommand{\abraket}[2]{\big< #1 \vphantom{#2} \big| \big. #2 \vphantom{#1} \big>} 
\newcommand{\Time}{\ensuremath{t}}

\newcommand{\Energy}{\ensuremath{E}}

\newcommand{\DeltaTime}{\ensuremath{\Delta t}}

\newcommand{\OrthogonalisationTime}{\ensuremath{t_\perp}}

\newcommand{\TimeInitial}{\ensuremath{t'}}
\newcommand{\TimeFinal}{\ensuremath{t''}}
\newcommand{\PositionInitial}{\ensuremath{x'}}
\newcommand{\PositionFinal}{\ensuremath{x''}}
\newcommand{\TimeFuture}{\ensuremath{t^+}}
\newcommand{\TimePast}{\ensuremath{t^-}}
\newcommand{\PositionFuture}{\ensuremath{x^+}}
\newcommand{\PositionPast}{\ensuremath{x^-}}

\newcommand{\MouthFuture}{\ensuremath{\mathcal{W}^+}}
\newcommand{\MouthPast}{\ensuremath{\mathcal{W}^-}}

\newcommand{\CR}{\mathrm{CR}}
\newcommand{\CV}{\mathrm{CV}}
\newcommand{\CRState}{\ensuremath{\sigma}}
\newcommand{\CVState}{\ensuremath{\theta}}
\newcommand{\ClassI}{\mathrm{\ensuremath{\RomanNumeralCaps{1}}}}
\newcommand{\ClassII}{\mathrm{\ensuremath{\RomanNumeralCaps{2}}}}
\newcommand{\ClassIII}{\mathrm{\ensuremath{\RomanNumeralCaps{3}}}}
\newcommand{\ClassIIIA}{\mathrm{\ensuremath{\RomanNumeralCaps{3}\text{-}A}}}
\newcommand{\ClassIIIB}{\mathrm{\ensuremath{\RomanNumeralCaps{3}\text{-}B}}}

\newcommand{\DCTCsCRMap}{\ensuremath{D}}
\newcommand{\DCTCsMixing}{\ensuremath{d}}
\newcommand{\DCTCsCVMap}{\ensuremath{T}}
\newcommand{\PCTCsMap}{\ensuremath{P}}

\newcommand{\NumberLevels}{\ensuremath{N}}
\newcommand{\NumberModes}{\ensuremath{M}}

\newcommand{\Unitary}{\ensuremath{U}}
\newcommand{\Identity}{\mathds{1}}
\newcommand{\Rotation}{\ensuremath{R}}
\newcommand{\Swap}{\ensuremath{S}}

\newcommand{\UnitaryMultimode}{\ensuremath{U}}
\newcommand{\UnitaryTraced}{\ensuremath{W}}
\newcommand{\UnitaryTime}{\ensuremath{R}}
\newcommand{\qclock}{\ensuremath{\phi}}
\newcommand{\qClock}{\ensuremath{\Phi}}
\newcommand{\Weight}{\ensuremath{c}}
\newcommand{\Free}{\ensuremath{g}}
\newcommand{\FreeScaled}{\ensuremath{q}}
\newcommand{\Incomplete}{\ensuremath{h}}
\newcommand{\Power}{\ensuremath{p}}
\newcommand{\ProbabilisticFalse}{\ensuremath{\alpha}}
\newcommand{\ProbabilisticTrue}{\ensuremath{\beta}}
\newcommand{\ProbabilisticTrueScaled}{\ensuremath{r}}

\newcommand{\DeltaEnergy}{\ensuremath{\Delta E}}
\newcommand{\MapBracket}[1]{\big[ #1 \big]}
\newcommand{\MapBrackets}[2]{\big[ #1 \vphantom{#2} , #2 \vphantom{#1} \big]}

\newcommand{\abs}[1]{\left| #1 \right|}

\newcommand{\tr}{\mathrm{tr}}
\renewcommand{\v}[1]{\ensuremath{\mathbf{#1}}} 
\newcommand{\gv}[1]{\ensuremath{\bm{#1}}} 

\begin{document}
	
\title{Billiard-ball paradox for a quantum wave packet}

\author{Lachlan G. Bishop}
\email{lachlan.bishop@uq.net.au}
\author{Timothy C. Ralph}
\author{Fabio Costa}
\affiliation{%
	School of Mathematics and Physics, The University of Queensland, St. Lucia, Queensland 4072, Australia
}%

\date{\today}

\begin{abstract}
	Past studies of the billiard-ball paradox, a problem involving an object that travels back in time along a closed timelike curve (CTC), typically concern themselves with entirely classical histories, whereby any trajectorial effects associated with quantum mechanics cannot manifest. Here we develop a quantum version of the paradox, wherein a (semiclassical) wave packet evolves through a region containing a wormhole time machine. This is accomplished by mapping all relevant paths on to a quantum circuit, in which the distinction of the various paths is facilitated by representing the billiard particle with a clock state. For this model, we find that Deutsch's prescription (D-CTCs) provides self-consistent solutions in the form of a mixed state composed of terms which represent every possible configuration of the particle's evolution through the circuit. In the equivalent circuit picture (ECP), this reduces to a binomial distribution in the number of loops of time machine. The postselected teleportation prescription (P-CTCs) on the other hand predicts a pure-state solution in which the loop counts have binomial coefficient weights. We then discuss the model in the continuum limit, with a particular focus on the various methods one may employ in order to guarantee convergence in the average number of clock evolutions. Specifically, for D-CTCs, we find that it is necessary to regularise the theory's parameters, while P-CTCs alternatively require more contrived modification.
\end{abstract}

\maketitle

\section{Introduction}\label{sec:introduction}

One of the most perplexing predictions of the general theory of relativity is that of closed timelike curves (CTCs) \cite{van_stockum_gravitational_1938,godel_example_1949,tipler_rotating_1974,morris_wormholes_1988,alcubierre_warp_1994,everett_warp_1996,fermi_time_2018,mallary_closed_2018,ralph_spinning_2020}. Such objects are significant because they facilitate the means by which an object can travel to its own past. A consequence of this is that the same object is able to interact with its past self, and so the possibility for CTCs to exist raises questions regarding their expected compliance with causality. These exotic objects therefore provide compelling motivation for research into chronology-violating physical systems, or those which exhibit backwards time travel.

The potential for the manifestation of CTCs in our universe leads to issues regarding the standard laws of physics and their compatibility with time travel. At the heart of these lie time-travel paradoxes, wherein simple physical scenarios often give rise to apparent logical contradictions. To rectify such problems, one typically determines if any solutions can be formulated that are void of pathological inconsistencies. The defining condition for such a methodology is formally encapsulated by Novikov's \emph{principle of self-consistency} \cite{novikov_analysis_1989-1}. Under this, only the solutions that do not possess paradoxical histories are considered to be physically meaningful.

An exemplary scenario of a time-travel paradox is known commonly as the \emph{billiard-ball paradox}. In this problem, a solid, spherical mass (the ``billiard ball'') enters a time machine, travels back to its own past and subsequently collides with its younger self. Given the initial position and velocity of the ball, studies \cite{friedman_cauchy_1990,echeverria_billiard_1991,lossev_jinn_1992,novikov_time_1992,mikheeva_inelastic_1993,mensky_three-dimensional_1996,dolansky_billiard_2010} have shown that there can be multiple self-consistent solutions (trajectories) which satisfy the equations of motion. This is of course in stark contrast with the determinism traditionally associated with classical mechanics, and so given its fundamentally indeterministic nature, quantum mechanics was proposed as a solution \cite{friedman_cauchy_1990}.

In a previous paper \cite{bishop_time-traveling_2021}, considering a quantum billiard ball on a classical trajectory, we showed how the problem of indeterminacy could indeed be quantum mechanically resolved in distinct ways by the two major quantum theories of CTCs: the Deutsch model (D-CTCs) \cite{deutsch_quantum_1991} and postselected teleportation (P-CTCs) \cite{lloyd_quantum_2011,lloyd_closed_2011}. The time-travelling particle model employed in this study was however limited in scope, as it was restricted to a set of highly idealised paths through the time-machine region. Here, we directly address this by extending our treatment to consider characteristically quantum trajectories.

Our interest therefore lies in the question of whether this complexification --- the inclusion of more exotic, quantum histories --- constitutes a model that exhibits rich physics while still admitting self-consistent solutions. Our findings suggest that this is true: for D-CTCs, we obtain a set of mixed states, and for P-CTCs, we obtain a pure state. Both of these results share a resemblance in the sense that they each display a distinct form of binomial distribution. This manifests in the P-CTC solution as binomial coefficients, expressing the multiplicity in the ways in which the particle can complete a certain number of loops of the time machine, attached to the spectrum of loop-counting ``clock'' states. The D-CTC result is similar, except that it is a mixture with a continuously parametrised structure inherited from the theory's nonuniqueness.

Perhaps the most significant regime of our model is the continuum limit, as this is the case which is the most physically meaningful. It is therefore intriguing that we should discover neither D-CTCs nor P-CTCs, without modification, provide solutions in this regime that express a finite average number of loops. In other words, when we move to the continuum, the particle becomes trapped in the time machine, which is at odds with any expectation of an ordinary evolution. Our answer to this is to regularise the theories. For D-CTCs, this can be accomplished simply by rescaling the model's inherent free parameters such that the maximum number of possible loops of the time machine scales directly with the probability of the loops to be vacuous, i.e., the mode is unoccupied. By contrast, P-CTCs require a more artificial form of regularisation, and we show how this may be accomplished in two distinct ways: by either adjusting both the postselected and prepared states away from maximal entanglement, or by introducing a finite chance that the scattering interaction does not occur.

We begin with a review of both models of quantum time travel in Sec.~\ref{sec:prescriptions}, followed by a discussion of the billiard-ball paradox, including indications of the way in which we extend the problem, in Sec.~\ref{sec:billiard-ball_paradox}. We then describe our quantum billiard-ball paradox model in detail in Sec.~\ref{sec:model}, starting with an introduction to the quantum clock states we use (Sec.~\ref{sec:clocks}), and ending with the specification of the simple $2$-mode version of the model (Sec.~\ref{sec:model_2}). The general $\NumberModes$-mode model is subsequently introduced in Sec.~\ref{sec:results}, in which results for both of D-CTCs and P-CTCs are formulated in Secs.~\ref{sec:results_D-CTCs} and \ref{sec:results_P-CTCs} respectively. This is then followed in Sec.~\ref{sec:limits} by analysis and discussion of the model in the continuum limit, for which a few distinct methods of ensuring well-behaved measurement probabilities are examined. We discuss a few more issues of the model in Sec.~\ref{sec:other}, and finish with concluding remarks in Sec.~\ref{sec:conclusion}.

\section{Background}\label{sec:background}

\subsection{Quantum time travel}\label{sec:prescriptions}

Given the discovery and subsequent investigation of CTCs in a semiclassical context, it is natural to seek theories of time travel that are fully compatible with quantum mechanics. Short of an accepted quantum theory of gravity, such formulations serve to deepen our understanding of time travel as a whole, and so provide a stimulating topic for study. Of all the research accomplished so far, two theories stand prominent above the rest. The first of these is the \emph{Deutsch model} (D-CTCs) \cite{deutsch_quantum_1991}, wherein the self-consistency principle is applied directly to the density matrix. The second is \emph{postselected teleportation} (P-CTCs) \cite{lloyd_quantum_2011,lloyd_closed_2011}, which provides self-consistent resolutions to temporal paradoxes through the use of a quantum communication channel formed by quantum teleportation in conjunction with postselection. Note that we will not concern ourselves with any alternative prescriptions of quantum time travel \cite{greenberger_quantum_2005, allen_treating_2014, araujo_quantum_2017, czachor_time_2019, baumeler_reversible_2019, tobar_reversible_2020} in the paper, owing to their lesser development in the literature.

First we shall consider D-CTCs, which are based on the requirement that time evolution of the local density operator (taken as a fundamental object) of a quantum system must be self-consistent. For this, we define two states: a chronology-respecting (CR) system $\op{\CRState}\in\mathcal{L}(\mathcal{H}_\CR)$ and a chronology-violating (CV) one $\op{\CVState}\in\mathcal{L}(\mathcal{H}_\CV)$, where $\mathcal{L}(\mathcal{H})$ denotes the space of linear operators on a Hilbert space $\mathcal{H}$. The former of these states describes an external system inbound to the CTC on which the latter state is trapped. The interaction between the two is described by the unitary $\op{\Unitary}$, after which the output subsystems are found via the maps
\begin{subequations}
	\begin{align}
		\map{\DCTCsCRMap}_{\Unitary}\MapBrackets{\op{\CRState}}{\op{\CVState}} &\equiv \tr_\CV\left[\op{\Unitary}\left(\op{\CRState} \otimes \op{\CVState}\right)\op{\Unitary}^\dagger\right] \in \mathcal{L}(\mathcal{H}_\CR), \label{eq:Deutsch_CR} \\
		\map{\DCTCsCVMap}_{\Unitary}\MapBrackets{\op{\CRState}}{\op{\CVState}} &\equiv \tr_\CR\left[\op{\Unitary}\left(\op{\CRState} \otimes \op{\CVState}\right)\op{\Unitary}^\dagger\right] \in \mathcal{L}(\mathcal{H}_\CV). \label{eq:Deutsch_CV}
	\end{align}
\end{subequations}
Deutsch's prescription then recovers self-consistent solutions by demanding that solution(s) of the CTC state be fixed point(s) of the CV map (\ref{eq:Deutsch_CV}), i.e.,
\begin{equation}
	\op{\CVState} = \map{\DCTCsCVMap}_{\Unitary}\MapBrackets{\op{\CRState}}{\op{\CVState}}. \label{eq:Deutsch_CV_2}
\end{equation}
Any such solutions can then be used to determine the output CR state in accordance with (\ref{eq:Deutsch_CR}).

P-CTCs, in contrast, are markedly simpler. One simply defines the reduced operator $\op{\UnitaryTraced} \equiv \tr_\CV\left[\op{\Unitary}\right]$, and uses it to compute the evolution of the pure system input state $\op{\CRState} = \aket{\psi}\abra{\psi}$ as
\begin{equation}
	\map{\PCTCsMap}_{\Unitary}\MapBracket{\op{\CRState}} = \frac{\op{\UnitaryTraced} \op{\CRState} \op{\UnitaryTraced}^\dagger}{\tr\left[\op{\UnitaryTraced} \op{\CRState} \op{\UnitaryTraced}^\dagger\right]} \sim \op{\UnitaryTraced} \aket{\psi}. \label{eq:P-CTCs_map}
\end{equation}
Here, the relation operator $\sim$ indicates the non-normalised pure state form of the preceding expression.

Both prescriptions are evidently major departures from standard quantum mechanics, but necessarily must be so in order to fulfil the seemingly rigid requirement of self-consistency in their respective quantum senses. It is important to note that for each model, self-consistency is achieved in a fundamentally distinct way. For D-CTCs, given that the fixed-point condition (\ref{eq:Deutsch_CV_2}) pertains to the density matrix as opposed to the individual states (in a pure-state decomposition sense), the Deutsch model intrinsically treats density matrices as ontic, ``real'' objects rather than as epistemic state representations. The resulting nonlinearity is therefore unique to the model. P-CTCs on the other hand are fundamentally equivalent to a path-integral formulation \cite{politzer_path_1994}, which means that the theory's inherent nonlinearity in state evolution is, like in the case of D-CTCs, a stark departure from ordinary quantum mechanics.

\subsection{A quantum billiard-ball paradox}\label{sec:billiard-ball_paradox}

Given any quantum theory of time travel (such as D-CTCs or P-CTCs), it is straightforward to analyse classical billiard-ball paradoxes from a quantum perspective. See, for example, our previous work \cite{bishop_time-traveling_2021}, wherein we studied a quantum circuit model of the classical trajectories of a simple billiard-ball paradox scenario. Here, we were able to assign probabilities to the set of paths through the time machine region, thereby rectifying the classical issue of indeterminism. However, despite the particle being treated quantum mechanically, the trajectories upon which it evolved were entirely classical, which is an aspect of the study that constitutes a major physical limitation. A significantly more realistic problem would be one in which quantum evolutions, i.e., those characterised by indeterminate simultaneous spatial and temporal localisation of the particle, are employed.

Although there do exist some early efforts to investigate such phenomena (see \cite{thorne_laws_1991} for an overview), most past studies of time travel paradoxes which contain interacting physical systems \cite{friedman_cauchy_1990,echeverria_billiard_1991,lossev_jinn_1992,novikov_time_1992,mikheeva_inelastic_1993,mensky_three-dimensional_1996,dolansky_billiard_2010} consider only those with characteristically classical histories. This is likely due to quantum evolutions being inherently more difficult to characterise and describe, especially in a continuous-variable context. It is therefore natural to wonder how a fully quantum billiard-ball paradox might be formulated and subsequently resolved.

Forms of the billiard-ball paradox posed in the aforementioned past literature typically involve (at least) two spatial dimensions. This allows the various trajectories of the time-travelling object to manifest with a wide diversity of complexification (multiple loops, varying angles and velocities, etc.). For simplicity, the seminal work of Friedman \emph{et al.} \cite{friedman_cauchy_1990} on the billiard-ball paradox and our later quantum extension \cite{bishop_time-traveling_2021} both consider an archetypal form of the paradox consisting of only two paths through the time machine region. By restricting the scope of the problem to this set of trajectories, the billiard-ball paradox is effectively limited to a $(1+1)$-dimensional problem (albeit with at most two nondegenerate spatial dimensions).

In such a simplified problem, given elastic collisions, the only interaction that can occur is a complete exchange of momentum. If we go one step further and reduce this problem to its simplest form (with a single spatial dimension), the only additional requirement that is needed involves the dynamical time-delayed wormhole, i.e., one that materialises and vanishes at precisely the correct times so as to produce appropriate CTCs. In this case, given that the billiard ball is confined to the two relevant trajectories, the single spatial dimension is no longer a meaningful degree of freedom and so it can be naturally abstracted away in the quantum circuit model.

\begin{figure*}[t]
	\includegraphics[scale=1.35]{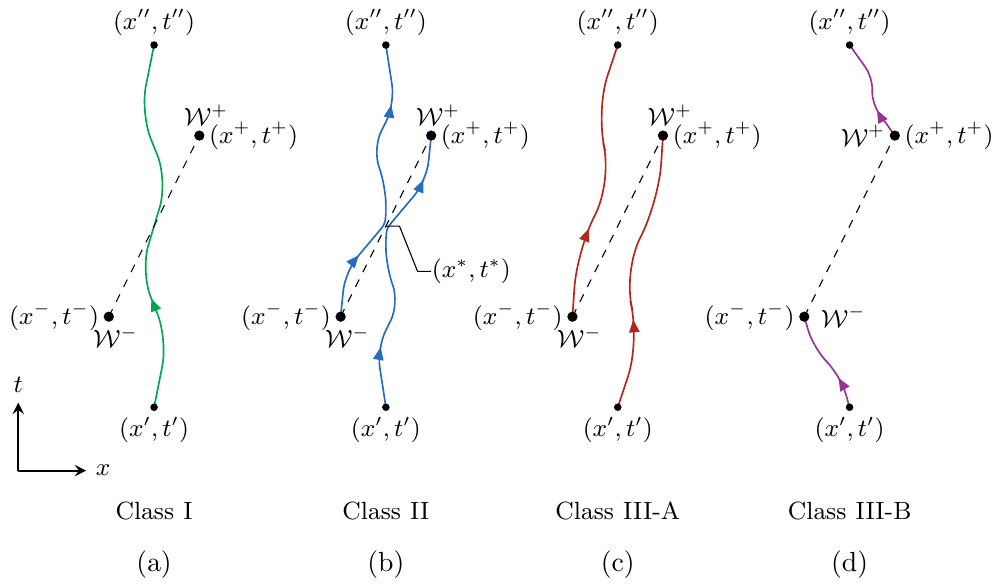}
	\caption[Evolution classes in a time-machine region]{\label{fig:evolution_classes}Examples of the classes of quantum trajectories of a wave packet through a CTC-wormhole spacetime. The initial coordinates are $(\PositionInitial,\TimeInitial)$ and the final coordinates are $(\PositionFinal,\TimeFinal)$, while the wormhole's past and future mouths $\MouthPast$ and $\MouthFuture$ are located at $(\PositionPast,\TimePast)$ and $(\PositionFuture,\TimeFuture)$ respectively. (a) depicts a chronology-respecting history (evolution class $\ClassI$) in which the wave packet evolves freely without entering the wormhole. (b) shows a chronology-violating history (evolution class $\ClassII$) in which the wave packet is scattered onto the CTC path by its future self, thereby causing it time travel into its past and subsequently strike its past self onto the CTC. (c) and (d) illustrate the two different ways (classes $\ClassIIIA$ and $\ClassIIIB$) in which the wormhole is traversed but no scattering self-interaction occurs. For all evolutions, the initial and final conditions are exactly the same, meaning that the histories are classically indistinguishable.}
\end{figure*}

In this work, we essentially reintroduce spatial and temporal degrees of freedom back into the model. For the propagation of a time-travelling quantum particle, the inclusion of such degrees of freedom gives rise to a more general version of billiard-ball paradox --- one which contains a spectrum of histories more exotic than its classical counterpart. Perhaps the simplest prototypical form of such a scenario is depicted figure \ref{fig:evolution_classes}. Here, the main classes of quantum trajectories which a wave packet may take through a CTC-wormhole region are illustrated, with classes $\ClassI$ and $\ClassII$ being those whose classical counterparts form the basis for most past work in the literature.

Due to its more realistic resemblance of actual physical trajectories, a model of the billiard-ball paradox with characteristically quantum evolutions would therefore make for a compelling study. To accomplish exactly this, we develop such a model, one which considers the simplest quantum case: a time-travelling wave packet propagating through a chronology-violating region in the fashion of figure \ref{fig:evolution_classes}. Our work is based upon a few main premises:
\begin{enumerate}[label=(\roman*),itemsep=0.015cm]
	\item By considering a simplified $(1+1)$-dimensional version of the problem, we restrict our attention to trajectories of classes $\ClassI$ and $\ClassII$ (as per figure \ref{fig:evolution_classes}). Physically, this is founded on the assumption that the wave packet is initially sufficiently narrow such that the contribution from trajectories of class $\ClassIII$ is negligible.
	\item We model the spatial and temporal localisation of the wave packet as a single-particle quantum state entangled across multiple quantum modes. These modes are in essence merely the aforementioned reintroduced degrees of freedom, and as such are taken to describe distinct localisations in spacetime.
\end{enumerate}
In order to facilitate the desired evolution of the quantum state and provide a method with which the various paths can be distinguished, we employ the additional techniques introduced in the previous work \cite{bishop_time-traveling_2021}:
\begin{enumerate}[label=(\roman*),itemsep=0.015cm]
	\setcounter{enumi}{2}
	\item We map the problem to a quantum circuit representation, and in doing so we use a vacuum state to allow the particle to be present or absent from particular paths.
	\item We attach an internal degree of freedom, one which operates like a clock, to the particle's quantum description, so that a measurement of this degree of freedom's final state can reveal the trajectory that was realised.
\end{enumerate}

The overarching idea is that our model simulates the elastic scattering of a quantum particle with its past or future self. Importantly, the spatial and temporal indeterminism of the particle's trajectories and interaction are captured by the use of a multimode clock-vacuum entangled state. An interpretation of this system is that it describes a localised quantum wave packet instead of a classical particle (for which position and momentum are simultaneously determinable with arbitrary precision), which greatly advances the realism of our earlier model \cite{bishop_time-traveling_2021}. We show how this enrichment leads to various physical consequences, such as multiple interactions inside the chronology-violating region (and therefore multiple traversals of the wormhole). Our findings reveal that, while the main models of quantum time travel, Deutsch's prescription (D-CTCs) and postselected teleportation (P-CTCs), both manage to provide self-consistent solutions, their structures are considerably distinct. For D-CTCs, the classical solution multiplicity is reproduced in the form of a set of mixed states, while P-CTCs predict an equal superposition of all possible trajectories through the time-machine region. We also find that, in the continuum limit of our model, both prescriptions only produce meaningful predictions if given particular regulations.

\section{Model}\label{sec:model}

\subsection{Billiard-ball paradox using discretely localised wave packets}

As we shall see, a quantum circuit formulation of a self-interacting time-travelling wave packet provides a useful tool with which we investigate some of the more general quantum characteristics of the corresponding semiclassical problem. In the most basic case, we can model the spatial localisation of such a wave packet as a quantum state which is entangled across multiple quantum subsystems. These ``modes'' are prescribed to correspond to distinct spatial localisations of the particle; a mode being occupied (i.e., containing a nonvacuous energy state) indicates the presence of the particle at that location. The idea is to then devise a quantum circuit that uses this multimode approach to model a more realistic version of the billiard-ball paradox --- one that considers characteristically quantum evolutions.

\begin{figure*}
	\includegraphics[scale=2]{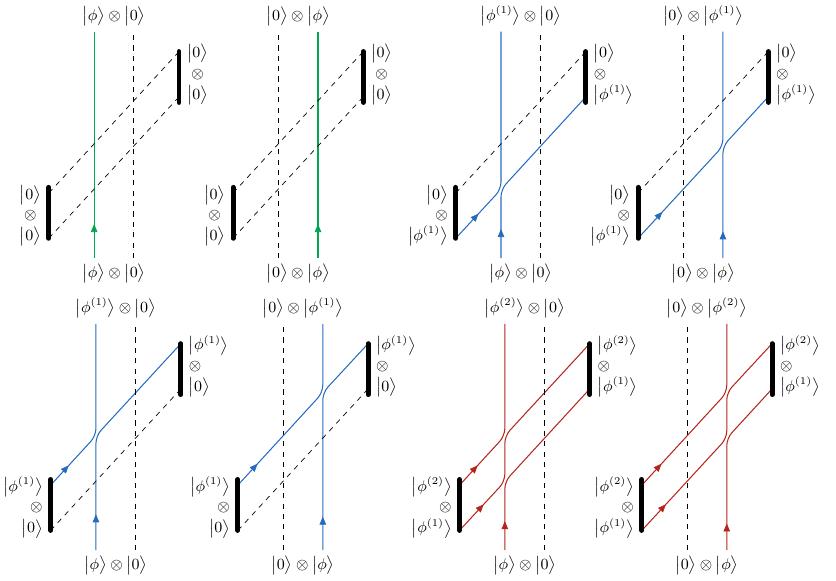}
	\caption[Trajectories in a noninfinitesimal time-machine region]{\label{fig:billiard_clock_evolutions}The complete set of self-consistent trajectories of a delocalised quantum particle (denoted by the state $\aket{\qclock}$) through $(1+1)$-dimensional spacetime containing a temporally extended CTC-wormhole time machine (the mouths are the very thick, solid lines). The modes represent spatial localisation of the particle; a solid, oriented line indicates the presence and direction of the particle while a dotted line indicates absence (i.e., vacuum $\aket{0}$). The bracketed superscripts attached to the particle states indicate the number of times that it has self-interacted (and therefore looped through the wormhole). Familiar quantum notation is used here to suggestively foreshadow the corresponding quantum circuit model.}
\end{figure*}

It is perhaps easiest to comprehend this notion by exploring the simplest case. This involves a $2$-mode (spatially) localised wave packet which propagates through a $(1+1)$-dimensional time-machine region containing a CTC-wormhole of $2$-mode (temporal) localisation (i.e., the mouths exists for a small but finite time). While the self-scattering of the particle remains entirely elastic, its spatial uncertainty means that any scattering which occurs is necessarily indeterministic in location. In other words, the particle may be kicked into the wormhole at two locations: either ``left'' (closer to the past mouth $\MouthPast$) or ``right'' (closer to the future mouth $\MouthFuture$) depending on its initial mode occupation. Additionally, given the wormhole's noninfinitesimal mouth duration, the particle may also collide with its past or future self ``early'' or ``late'' (or indeed both) on its evolution through the region. This is illustrated in $(1+1)$-dimensions in figure \ref{fig:billiard_clock_evolutions}.

Here, the particle's internal state is denoted as $\aket{\qclock}$, while the state $\aket{0}$ represents the physical absence of a particle (hence the ``vacuum'' nomenclature). The entanglement of the particle $\aket{\qclock}$ with the vacuum $\aket{0}$ over the two modes means that the state
\begin{equation}
	\aket{\qClock_2} = \sqrt{\Weight_1}\aket{\qclock}\otimes\aket{0} + \sqrt{\Weight_2}\aket{0}\otimes\aket{\qclock} \label{eq:entangled_2_pure}
\end{equation}
represents a particle that is in a superposition of localisations (with weights $\Weight_1$ and $\Weight_2$ respectively). Given that the wormhole mouths exist for a long enough time such that two temporally distinct CTCs manifest, then we have at most eight self-consistent trajectories through the region, with two of these containing two self-interactions as represented in figure \ref{fig:billiard_clock_evolutions}.

\subsection{Quantum clocks}\label{sec:clocks}

In order to lend operational distinction to the relatively large set of possible evolutions the particle may take through the chronology-violating region, we shall introduce an internal degree of freedom to the particle in the form of a quantum clock. The basic function of such an object is to, through measurement, tell us how much time has passed in its reference frame. This in turn allows us to distinguish between the distinct trajectories through the chronology-violating region.

A quantum clock is any state $\aket{\qclock}$ which is definable in tandem with the time evolution operator $\op{\UnitaryTime}(\DeltaTime)$ as a one-parameter state that obeys the action
\begin{equation}
	\aket{\qclock(\Time+\DeltaTime)} = \op{\UnitaryTime}(\DeltaTime)\aket{\qclock(\Time)}. \label{eq:clock_evolution}
\end{equation}
Given this characterisation, the lone parameter (such as $\Time$ in $\aket{\qclock(\Time)}$) is interpretable as the ``time'' of the clock, and corresponds physically to an internal degree of freedom in the associated quantum system. For our purposes, we attach such a clock state to the particle's quantum description, which results in the parameter being a measure of time experienced by the particle in its own reference frame. Without loss of generality, such a state can be expressed at an arbitrary time $\Time$ in terms of an $\NumberLevels$-dimensional set of orthonormal energy states $\left\{\aket{n}\right\}_{n=1}^\NumberLevels$ which collectively form a basis for the clock's Hilbert space. The most general evolution in this energy basis is the construction
\begin{equation}
	\aket{\qclock(\Time)} = \frac{1}{\sqrt{\NumberLevels}}\sum_{n=1}^{\NumberLevels}\e^{-\eye \Energy_n \Time/\hbar}\aket{n}, \label{eq:clock_evolved}
\end{equation}
where $\Energy_n$ is the energy of the $n$th eigenstate $\aket{n}$. The main function we require of these clocks is to produce a useful measure of similarity between any two states, like $\aket{\qclock(\Time)}$ and $\aket{\qclock(\Time + \DeltaTime)}$, via an overlap measurement, e.g., $\abraket{\qclock(\Time)}{\qclock(\Time+\DeltaTime)}$. The simplest way to accomplish this is to require that the energies be equally spaced such that
\begin{equation}
	\Energy_n = \Energy_1 + (n-1)\DeltaEnergy. \label{eq:clock_energy}
\end{equation}
We may then define the quantity
\begin{equation}
	\OrthogonalisationTime = \frac{2\pi\hbar}{\NumberLevels\DeltaEnergy}, \label{eq:orthogonalisation_time}
\end{equation}
called the orthogonalisation time, which quantifies the smallest unit of time an $\NumberLevels$-level clock state requires in order to evolve into an orthogonal state. Its use is revealed when we use it alongside (\ref{eq:clock_evolved}) to write the overlap measurement as
\begin{align}
	\abraket{\qclock(\Time)}{\qclock(\Time+\DeltaTime)} = \frac{\e^{-\eye E_1\DeltaTime/\hbar}}{\NumberLevels} \sum_{n=1}^{\NumberLevels} \exp\left[-2\pi\eye\frac{n-1}{\NumberLevels}\frac{\DeltaTime}{\OrthogonalisationTime}\right]. \label{eq:clock_overlap_orthogonalisation}
\end{align}
Evidently, in the case that any measured evolution time difference is equal to the orthogonalisation time, i.e, $\DeltaTime = \OrthogonalisationTime$, then one can show that the clock overlap vanishes, i.e.,
\begin{equation}
	\left.\abraket{\qclock(\Time)}{\qclock(\Time+\DeltaTime)}\right|_{\DeltaTime = \OrthogonalisationTime} = 0.
\end{equation}
In other words, $\OrthogonalisationTime$ quantifies the smallest amount of time required for a clock to evolve into an orthogonal state. This then defines a ``resolution'' of the clock: the smallest interval of time it can distinguish. The $\OrthogonalisationTime$-duration periodicity of the overlap function (\ref{eq:clock_overlap_orthogonalisation}) means that there exists $\NumberLevels$ mutually orthogonal clock states $\left\{\op{\UnitaryTime}^{n}(\OrthogonalisationTime)\aket{\qclock}\right\}_{n=0}^{\NumberLevels-1}$. This is a useful feature, as it allows one to distinguish between multiple loops of the CTCs, provided the clock's orthogonalisation time is set such that it exactly matches the wormhole time delay.

\subsection{Two-mode quantum circuit model of the billiard-ball paradox}\label{sec:model_2}

\begin{figure}
	\hspace*{-0.5cm}\includegraphics{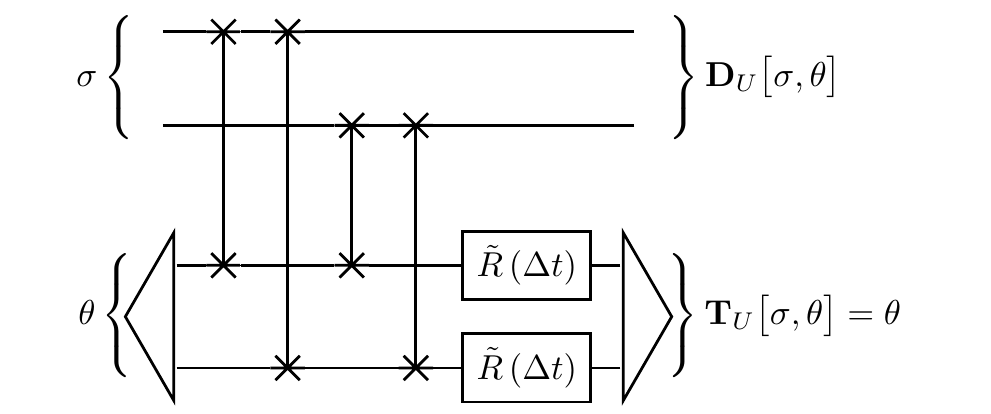}
	\caption[Two-mode quantum circuit of a time-travelling quantum state]{\label{fig:circuit_two-mode}The quantum circuit representation of the trajectories depicted in figure \ref{fig:billiard_clock_evolutions}. Here, the \textgate{swap} gates correspond to the ``extended'' \textgate{swap} as per Eq.~(\ref{eq:SWAP_vacuum}).}
\end{figure}

Given the clocks states introduced immediately prior, we can now map the set of trajectories for the $2$-mode case (figure \ref{fig:billiard_clock_evolutions}) directly to a quantum circuit model. Figure \ref{fig:circuit_two-mode} depicts the circuit, wherein the operation of the clock [i.e., time evolution as per Eq.~(\ref{eq:clock_evolution})] is incorporated by placing time evolution gates in the CTC modes. The duration $\DeltaTime$ therefore represents the CTC's time delay (the amount of time between the wormhole's mouths). One can check that the circuit does indeed operate in the desired manner by simply verifying that its action on all classically localised states coincides with the complete set of subfigures in figure \ref{fig:billiard_clock_evolutions}.

It is important to mention a couple of simplifications we have imposed in order to make the formulation of the model significantly more feasible. First is that we omit any external time evolution, such as that prior to or after the ``collision'' (emulated here by the sequence of \textgate{swap} gates). Second is that we assume no dispersion, meaning that there is no interaction between modes in either the chronology-respecting or chronology-violating bundles. This is thought to be a direct consequence of the principle of self-consistency, and we discuss this matter in Sec.~\ref{sec:dispersion}.

In accordance with our prescription of the wave packet's spatialisation, the input state for our circuit here is simply the density representation of (\ref{eq:entangled_2_pure}), i.e.,
\begin{equation}
	\op{\CRState} = \aket{\qClock_2}\abra{\qClock_2}. \label{eq:entangled_2_density}
\end{equation}
Time evolution of the clock on the CV modes, as per the action (\ref{eq:clock_evolution}), will be expressed in the form
\begin{equation}
	\op{\Rotation}^k(\DeltaTime)\aket{\qclock} \equiv \aket{\qclock^{(k)}(\DeltaTime)}
\end{equation}
so that the composite entanglement (which includes vacuum states) undergoes the evolution
\begin{equation}
	\op{\vac{\Rotation}}^k(\DeltaTime) \otimes \op{\vac{\Rotation}}^k(\DeltaTime) \aket{\qClock_2} \equiv \aket{\qClock_2^{(k)}(\DeltaTime)}
\end{equation}
via a vacuum-inclusive unitary
\begin{equation}
	\op{\vac{\Rotation}}(\DeltaTime) = \aket{0}\abra{0} + \op{\Rotation}(\DeltaTime).
\end{equation}
Note that here, we implicitly defined (without loss of generality) the zeroth evolution amplitude as
\begin{equation}
	\abra{0}\op{\vac{\Rotation}}\aket{0} = 1. \label{eq:rotation_vacuum_unity}
\end{equation}

In our quantum circuit framework, the (neighbouring-mode) \textgate{swap} gate
\begin{equation}
	\op{\Swap} = \sum_{i,j=1}^{\NumberLevels}{\aket{i}\abra{j}}\otimes{\aket{j}\abra{i}} \label{eq:SWAP}
\end{equation}
precisely emulates an elastic self-collision in the [$(1+1)$-dimensional] corresponding mechanical problem. (Note that here, the basis states correspond to those from which the particle's wave function is constructed. This means that, in the case of a quantum clock, these basis states represent the number states of the clock's Hilbert space.) This is because the complete transference of momentum between two particles (here, the past and future versions of the billiard ball) is necessarily an exchange of quantum clock state, provided the associated particles are otherwise identical. To clarify, in a particle's quantum description, the clock degree of freedom is the identifying attribute, while momentum performs the same function in its classical description. As such, the two necessarily correspond such that when a pair of classically indistinguishable particles trade momentum, they exchange their clocks, i.e., $\op{\Swap} \aket{\qclock_\textrm{A}} \otimes \aket{\qclock_\textrm{B}} = \aket{\qclock_\textrm{B}} \otimes \aket{\qclock_\textrm{A}}$.

For the model to function as intended, we further mandate that the vacuum and clock cannot ``collide'' with each other. To enforce this, the \textgate{swap} gate (written here for neighbouring modes for simplicity) must be modified to the form
\begin{align}
	\op{\vac{\Swap}} &= \op{\vac{\Identity}}\otimes\aket{0}\abra{0} + \aket{0}\abra{0}\otimes\op{\vac{\Identity}} \nonumber\\
	&\quad - \aket{0}\abra{0}\otimes\aket{0}\abra{0} + \op{\Swap} \label{eq:SWAP_vacuum}
\end{align}
where $\op{\vac{\Identity}} = \op{\Identity} + \aket{0}\abra{0}$ is the identity matrix in a vacuum-inclusive Hilbert space. This altered \textgate{swap} simply excludes the vacuum from swapping with the nonvacuous states, i.e.,
\begin{subequations}
\begin{align}
	\op{\vac{\Swap}} \aket{0} \otimes \aket{0} &= \aket{0} \otimes \aket{0}, \\
	\op{\vac{\Swap}} \aket{\qclock} \otimes \aket{0} &= \aket{\qclock} \otimes \aket{0}, \\
	\op{\vac{\Swap}} \aket{0} \otimes \aket{\qclock} &= \aket{0} \otimes \aket{\qclock}, \\
	\op{\vac{\Swap}} \aket{\qclock_\textrm{A}} \otimes \aket{\qclock_\textrm{B}} &= \aket{\qclock_\textrm{B}} \otimes \aket{\qclock_\textrm{A}}.
\end{align}
\end{subequations}
From these, one can subsequently construct the unitary corresponding to the $2$-mode circuit,
\begin{align}
	\op{\UnitaryMultimode}_2 &= \left(\op{\vac{\Identity}}_1 \otimes \op{\vac{\Identity}}_2 \otimes \op{\vac{\Rotation}}_3 \otimes \op{\vac{\Rotation}}_4\right) \op{\vac{\Swap}}_{2,4} \, \op{\vac{\Swap}}_{2,3} \, \op{\vac{\Swap}}_{1,4} \, \op{\vac{\Swap}}_{1,3} \label{eq:unitary_2}
\end{align}
where the modes $\{1,2\}$ and $\{3,4\}$ are the pairs of CR and CV modes respectively, and $\op{\vac{\Swap}}_{i,j}$ represents a vacuum-inclusive \textgate{swap} gate between the $i$th and $j$th modes [see Eq.~(\ref{eq:SWAP_multimode}) for a general form].

\subsubsection{Two-mode solutions}

With our complete description of the circuit, we now search for self-consistent solutions given the clock input state (\ref{eq:entangled_2_density}). In accordance with D-CTCs, appropriate solutions are found by first determining all fixed points of the trapped state. The most general form of this can be constructed using vacuum and clock states in each mode to define a basis of the state
\begin{align}
	\map{\DCTCsCVMap}_{\UnitaryMultimode_2} &= \Bigl(u\aket{0}\abra{0} + \sum_{n=0}^{\NumberLevels} a_{n,0}\aket{\qclock^{(n)}}\abra{0}\nonumber\\
	&\qquad+ \sum_{m=0}^{\NumberLevels} b_{0,m}\aket{0}\abra{\qclock^{(m)}} + \sum_{n,m=0}^{\NumberLevels} c_{n,m}\aket{\qclock^{(n)}}\abra{\qclock^{(m)}}\Bigr)\nonumber\\
	&\otimes \Bigl(v\aket{0}\abra{0} + \sum_{j=0}^{\NumberLevels} x_{j,0}\aket{\qclock^{(j)}}\abra{0}\nonumber\\
	&\qquad+ \sum_{k=0}^{\NumberLevels} y_{0,k}\aket{0}\abra{\qclock^{(k)}} + \sum_{j,k=0}^{\NumberLevels} z_{j,k}\aket{\qclock^{(j)}}\abra{\qclock^{(k)}}\Bigr),
	 \label{eq:D-CTCs_2_initial}
\end{align}
which accounts for all possible rotational ``times'' of the clock (given $\NumberLevels\geq2$). Here, $u$, $a_{n,0}$, $b_{0,m}$, $c_{n,m}$, $v$, $x_{j,0}$, $y_{0,k}$, and $z_{j,k}$ are complex parameters to be determined. Given that D-CTCs mix out clock states due to the trace operation, then we can immediately neglect terms with inappropriate mixing. In other words, any term which contains a vacuum-clock outer product (e.g., $\aket{\qclock^{(n)}}\abra{0}$ and $\aket{0}\abra{\qclock^{(n)}}$) vanishes when the CR modes are traced out, meaning that they can be neglected hereafter for simplification. Doing so yields the expression
\begin{align}
	\map{\DCTCsCVMap}_{\UnitaryMultimode_2} &= uv\aket{0}\abra{0}\otimes\aket{0}\abra{0} \nonumber\\
	&+ \sum_{n,m=0}^{\NumberLevels}c_{n,m}v\aket{\qclock^{(n)}}\abra{\qclock^{(m)}}\otimes\aket{0}\abra{0} \nonumber\\
	&+ \sum_{j,k=0}^{\NumberLevels}uz_{j,k}\aket{0}\abra{0}\otimes\aket{\qclock^{(j)}}\abra{\qclock^{(k)}} \nonumber\\
	&+ \sum_{n,m,j,k=0}^{\NumberLevels}c_{n,m}z_{j,k}\aket{\qclock^{(n)}}\abra{\qclock^{(m)}}\otimes\aket{\qclock^{(j)}}\abra{\qclock^{(k)}}.
\end{align}
Taking the tensor product of this form with the input density (\ref{eq:entangled_2_density}), evolving the composite result by the unitary (\ref{eq:unitary_2}), and tracing out the CR modes then gives
\begin{align}
	\map{\DCTCsCVMap}_{\UnitaryMultimode_2} &= uv\aket{0}\abra{0}\otimes\aket{0}\abra{0} \nonumber\\
	&+ \sum_{n,m=0}^{\NumberLevels}c_{n,m}v\aket{\qclock^{(1)}}\abra{\qclock^{(1)}}\otimes\aket{0}\abra{0} \nonumber\\
	&+ \sum_{j,k=0}^{\NumberLevels}uz_{j,k}\aket{0}\abra{0}\otimes\aket{\qclock^{(1)}}\abra{\qclock^{(1)}} \nonumber\\
	&+ \sum_{n,m,j,k=0}^{\NumberLevels}c_{n,m}z_{j,k}\aket{\qclock^{(1)}}\abra{\qclock^{(1)}}\otimes\aket{\qclock^{(1+n)}}\abra{\qclock^{(1+m)}}.
\end{align}
Matching coefficients with the initial form (\ref{eq:D-CTCs_2_initial}) as per the fixed-point condition (\ref{eq:Deutsch_CV_2}) subsequently reveals the sum
\begin{align}
	\map{\DCTCsCVMap}_{\UnitaryMultimode_2} &= \Free_{0,0} \aket{0}\abra{0} \otimes \aket{0}\abra{0} \nonumber\\
	&\quad + \Free_{1,0} \aket{\qclock^{(1)}}\abra{\qclock^{(1)}} \otimes \aket{0}\abra{0} \nonumber\\
	&\quad + \Free_{0,1} \aket{0}\abra{0} \otimes \aket{\qclock^{(1)}}\abra{\qclock^{(1)}} \nonumber\\
	&\quad + \Free_{1,1} \aket{\qclock^{(1)}}\abra{\qclock^{(1)}} \otimes \aket{\qclock^{(2)}}\abra{\qclock^{(2)}}. \label{eq:D-CTCs_CV_2}
\end{align}
where, for brevity, we assimilated the $a_{n,m}$ and $b_{j,k}$ products into single form $\Free_{u,v}$ coefficients. Individually, each term here is a solution in its own right, and so the linear combination of them constitutes the most general solution. Evidently, this trapped CV state represents the mixture of the possibilities of having no clock ($\aket{0}\abra{0} \otimes \aket{0}\abra{0}$), one clock in one of the modes ($\aket{\qclock^{(1)}}\abra{\qclock^{(1)}} \otimes \aket{0}\abra{0}$ and $\aket{0}\abra{0} \otimes \aket{\qclock^{(1)}}\abra{\qclock^{(1)}}$), and one clock in both ($\aket{\qclock^{(1)}}\abra{\qclock^{(1)}} \otimes \aket{\qclock^{(2)}}\abra{\qclock^{(2)}}$). Note that for the last term, the ``tick'' count of the clock in the second mode is one greater than in the first. Since the input is a single-particle state, then we can interpret the second-mode clock as being the trapped clock's future self.

Corresponding to this form, the CR output is then easily calculable to be
\begin{align}
	\map{\DCTCsCRMap}_{\UnitaryMultimode_2} &= \Free_{0,0} \aket{\qClock_2^{(0)}}\abra{\qClock_2^{(0)}} \nonumber\\
	&\quad + (\Free_{1,0} + \Free_{0,1}) \aket{\qClock_2^{(1)}}\abra{\qClock_2^{(1)}} \nonumber\\
	&\quad + \Free_{1,1} \aket{\qClock_2^{(2)}}\abra{\qClock_2^{(2)}}. \label{eq:D-CTCs_CR_2}
\end{align}
This mixed state tells us that the outgoing clock can have any number of ``ticks'' up to the limit set by the model (i.e., the number of modes in the bundle).

P-CTCs on the other hand are considerably more straightforward. First, one computes the reduced operator $\op{\UnitaryTraced}_2 = \tr_\CV[\op{\UnitaryMultimode}_2]$, with which it is easy to show that
\begin{align}
	\map{\PCTCsMap}_{\UnitaryMultimode_2}\MapBracket{\op{\CRState}}(\DeltaTime) &\sim \op{\UnitaryTraced}_2\aket{\qClock_2}\nonumber\\
	&= \aket{\qClock_2^{(0)}} + 2\aket{\qClock_2^{(1)}} + \aket{\qClock_2^{(2)}}. \label{eq:P-CTCs_2}
\end{align}
Recall that relation symbol $\sim$ indicates the non-normalised pure-state form of the preceding expression.

\section{Multimode quantum circuit model of the billiard-ball paradox}\label{sec:results}

Here, we generalise the basic $2$-mode circuit of the previous section to one of $\NumberModes$ modes, so as to produce the most general model of a time-travelling discretely localised wave packet. The input state $\CRState = \aket{\qClock_\NumberModes}\abra{\qClock_\NumberModes}$ is the entangled state
\begin{align}
	\aket{\qClock_\NumberModes} &= \sum_{m=1}^{\NumberModes}\sqrt{\Weight_m}\Bigl[\aket{0}^{\otimes(m-1)}\otimes\aket{\qclock}\otimes\aket{0}^{\otimes(\NumberModes-m)}\Bigr], \label{eq:clock_multimode}
\end{align}
which describes a single ordinary quantum clock state $\aket{\qclock}$ ``spread'' with the vacuum $\aket{0}$ over $\NumberModes$ modes, with the $m$th mode being weighted with amplitude $\sqrt{\Weight_m}$. In the first-quantised picture this is just a single particle in a superposition along $\NumberModes$ trajectories.

\begin{figure*}
	\hspace*{-1cm}\includegraphics[scale=1]{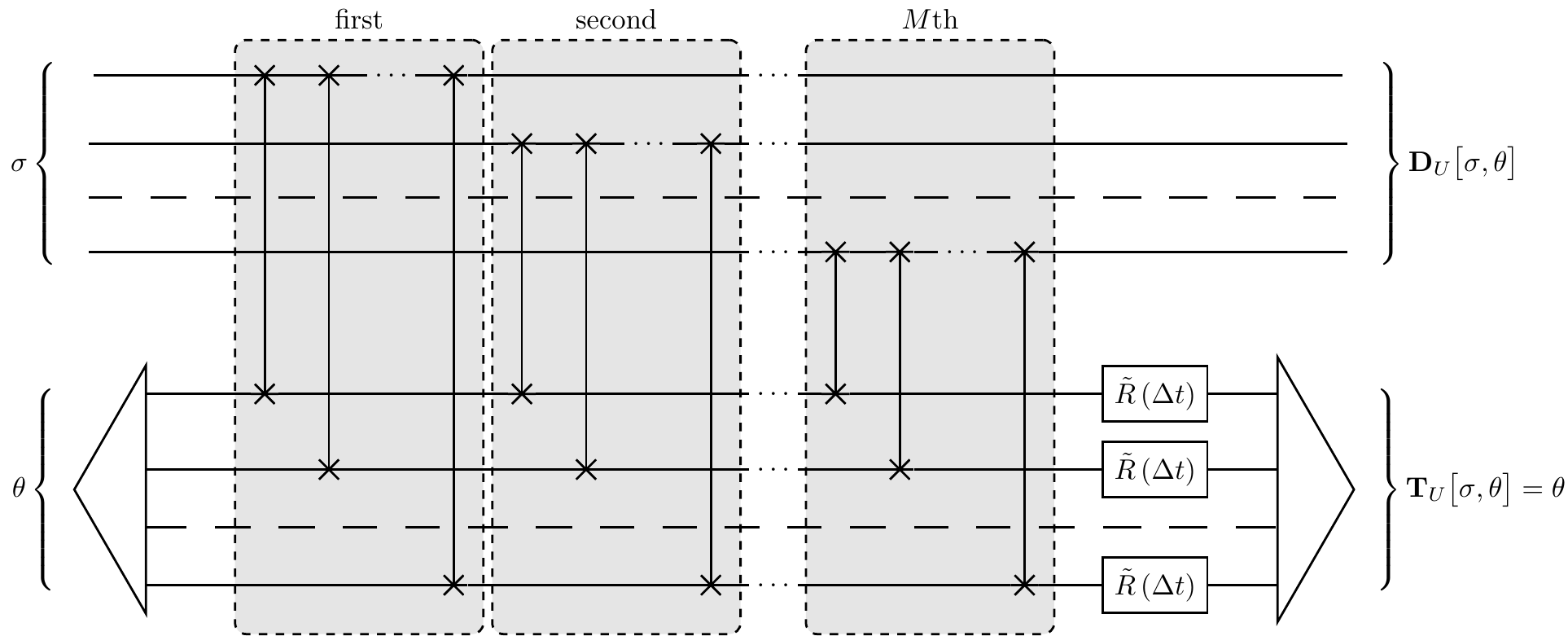}
	\caption[Quantum circuit of a time-travelling quantum state]{\label{fig:circuit_multimode}A multimode version of the $2$-mode circuit in figure \ref{fig:circuit_two-mode}. The dashed modes in both bundles represent the pictorial omission of $\NumberModes - 3$ modes between the second and $\NumberModes$th modes.}
\end{figure*}

Next, we define the general vacuum-modified \textgate{swap} gate as
\begin{align}
	\op{\vac{\Swap}}_{i,j} &= \op{\vac{\Identity}}^{\otimes(i-1)} \otimes \aket{0}\abra{0} \otimes \op{\vac{\Identity}}^{\otimes(2\NumberModes-i)} \nonumber\\
	&\quad + \op{\vac{\Identity}}^{\otimes(j-1)} \otimes \aket{0}\abra{0} \otimes \op{\vac{\Identity}}^{\otimes(2\NumberModes-j)} \nonumber\\
	&\quad - \op{\vac{\Identity}}^{\otimes(i-1)} \otimes \aket{0}\abra{0} \otimes \op{\vac{\Identity}}^{\otimes(j-i-1)} \otimes \aket{0}\abra{0} \otimes \op{\vac{\Identity}}^{\otimes(2\NumberModes-j)} \nonumber\\
	&\quad + \sum_{k,l = 1}^{\NumberLevels} \op{\vac{\Identity}}^{\otimes(i-1)} \otimes \aket{k}\abra{l} \otimes \op{\vac{\Identity}}^{\otimes(j-i-1)} \nonumber\\
	&\qquad \qquad \otimes \aket{l}\abra{k} \otimes \op{\vac{\Identity}}^{\otimes(2\NumberModes-j)}. \label{eq:SWAP_multimode}
\end{align}
The circuit unitary is then simply $\NumberModes$ sets of $\NumberModes$ \textgate{swap} operations, followed by the time evolutions of the CV modes,
\begin{align}
	\op{\UnitaryMultimode}_\NumberModes &= \left(\op{\vac{\Identity}}^{\otimes\NumberModes} \otimes \op{\vac{\Rotation}}^{\otimes\NumberModes}(\DeltaTime)\right) \nonumber\\
	&\quad \left(\op{\vac{\Swap}}_{\NumberModes,2\NumberModes} \ldots \op{\vac{\Swap}}_{\NumberModes,(\NumberModes+2)} \, \op{\vac{\Swap}}_{\NumberModes,(\NumberModes+1)}\right) \nonumber\\
	&\quad \ldots \left(\op{\vac{\Swap}}_{2,2\NumberModes} \ldots \op{\vac{\Swap}}_{2,(\NumberModes+2)} \, \op{\vac{\Swap}}_{2,(\NumberModes+1)}\right) \nonumber\\
	&\quad \left(\op{\vac{\Swap}}_{1,2\NumberModes} \ldots \op{\vac{\Swap}}_{1,(\NumberModes+2)} \, \op{\vac{\Swap}}_{1,(\NumberModes+1)}\right). \label{eq:unitary_multimode}
\end{align}
This is represented diagrammatically in figure \ref{fig:circuit_multimode}. It is important to note that here we choose the dimensionality of the clock (i.e., the number of orthonormal energy eigenstates in the clock's Hilbert space) $\NumberLevels$ to be larger than or equal to the number of modes $\NumberModes$. With the foresight that the clock evolves (``ticks'') up to $\NumberModes$ times in the circuit, this ensures that the clock has as many distinct (orthogonal) states as required, thereby allowing it to distinguish all possible rotational states.

Given the form of the input state (\ref{eq:clock_multimode}) and unitary (\ref{eq:unitary_multimode}), computing both D-CTC and P-CTC solutions is a highly nontrivial task. Fortunately, we can use the $2$-mode case of Sec.~\ref{sec:model_2} to form an intuitive understanding of how the prescriptions each provide self-consistent solutions. We will subsequently be able to construct generalised solutions to the $\NumberModes$-mode case.

\subsection{D-CTC solutions}\label{sec:results_D-CTCs}

For D-CTCs, we first desire to construct the general CV solution, which can be accomplished by considering the simpler $2$-mode case. Here, one has a total of $2^2 = 4$ independent fixed points, which collectively form the complete set of all possible combinations of clock and vacuum classical mixed products. The time evolution states of these clocks correspond to their ordered (left-to-right) count, e.g., $\aket{\qclock^{(1)}}\abra{\qclock^{(1)}} \otimes \aket{0}\abra{0}$ contains only a single clock and so its rotation is of single order, while $\aket{\qclock^{(1)}}\abra{\qclock^{(1)}} \otimes \aket{\qclock^{(2)}}\abra{\qclock^{(2)}}$ contains two clocks, and so the ``first'' clock must be of order one while the ``second'' clock must be of order two. This is to say that the binary-subscript coefficients $\Free_{u,v}$ in Eq.~(\ref{eq:D-CTCs_CR_2}) were purposefully designed such that a corresponding fixed point has $u$ clocks of order $u$ in the left mode and $v$ clocks of order $u+v$ in the right mode. Extending this idea to a general $\NumberModes$-mode case is therefore very straightforward. Considering free coefficients denoted as $\Free_{\alpha_1,\alpha_2,\ldots,\alpha_\NumberModes}$, one would find a corresponding fixed point
\begin{widetext}
\begin{align}
	\Free_{\alpha_1,\alpha_2,\ldots,\alpha_\NumberModes} \Longleftrightarrow&\left[\aket{0}\abra{0}\right]^{\otimes(1-\alpha_1)} \otimes \left[\aket{\qclock^{(\alpha_1)}}\abra{\qclock^{(\alpha_1)}}\right]^{\otimes\alpha_1} \nonumber\\
	&\otimes \left[\aket{0}\abra{0}\right]^{\otimes(1-\alpha_2)} \otimes \left[\aket{\qclock^{(\alpha_1+\alpha_2)}}\abra{\qclock^{(\alpha_1+\alpha_2)}}\right]^{\otimes\alpha_2} \nonumber\\
	&\otimes \ldots \otimes \left[\aket{0}\abra{0}\right]^{\otimes(1-\alpha_{\NumberModes})} \otimes \left[\aket{\qclock^{(\alpha_1+\alpha_2+\ldots+\alpha_\NumberModes)}}\abra{\qclock^{(\alpha_1+\alpha_2+\ldots+\alpha_\NumberModes)}}\right]^{\otimes\alpha_\NumberModes}. \label{eq:D-CTCs_CV_term}
\end{align}
\end{widetext}
The $\alpha_m$ subscripts are binary-valued, and essentially indicate the presence of a clock in the $m$th mode. For example, the term in the 3-mode case containing clocks in the first and last modes would appear uniquely as
\begin{equation}
	\Free_{1,0,1} \aket{\qclock^{(1)}}\abra{\qclock^{(1)}} \otimes \aket{0}\abra{0} \otimes \aket{\qclock^{(2)}}\abra{\qclock^{(2)}}.
\end{equation}
Consequently, there are $2^\NumberModes$ possible fixed points for $\NumberModes$ modes, each of which is a valid eigenstate of the CTC time evolution in its own right. Hereafter, these will be expressed using the simplifying index vector notation
\begin{equation}
	\gv{\alpha} \equiv \left(\alpha_1,\alpha_2,\ldots,\alpha_\NumberModes\right),
\end{equation}
so that $\abs{\gv{\alpha}} = \alpha_1+\alpha_2+\ldots+\alpha_\NumberModes$. Accordingly, the most general CV solution which can be constructed out of mixed terms of the form (\ref{eq:D-CTCs_CV_term}) is the convex combination
\begin{align}
	\map{\DCTCsCVMap}_{\UnitaryMultimode_\NumberModes} = \sum_{\gv{\alpha} = \v{0}}^{\v{1}} \Free_{\gv{\alpha}} &\left[\aket{0}\abra{0}\right]^{\otimes(1-\alpha_1)} \otimes \left[\aket{\qclock^{(\alpha_1)}}\abra{\qclock^{(\alpha_1)}}\right]^{\otimes\alpha_1} \nonumber\\
	&\otimes \left[\aket{0}\abra{0}\right]^{\otimes(1-\alpha_2)} \otimes \left[\aket{\qclock^{(\alpha_1+\alpha_2)}}\abra{\qclock^{(\alpha_1+\alpha_2)}}\right]^{\otimes\alpha_2} \nonumber\\
	&\otimes \ldots \otimes \left[\aket{0}\abra{0}\right]^{\otimes(1-\alpha_{\NumberModes})} \otimes \left[\aket{\qclock^{(\abs{\gv{\alpha}})}}\abra{\qclock^{(\abs{\gv{\alpha}})}}\right]^{\otimes\alpha_\NumberModes}. \label{eq:multimode_D-CTCs_CV_together}
\end{align}
After normalisation, we are left with $2^\NumberModes - 1$ nondegenerate, independent free variables $\left\{\Free_{\alpha_1,\alpha_2,\ldots,\alpha_\NumberModes}\right\}$ which collectively parametrise the entire mixture.

Given the term-wise correspondence of the $2$-mode CV (\ref{eq:D-CTCs_CV_2}) and CR (\ref{eq:D-CTCs_CR_2}) states highlighted by the $\Free_{u,v}$ coefficients, it is easy to see that the largest rotational clock order in a CV fixed point contributes directly to the entangled clock-vacuum state's rotational order in the CR solution. This is perhaps best represented mathematically,
\begin{widetext}
\begin{align}
	&\text{vacuum:} & &\map{\DCTCsCVMap}_{\UnitaryMultimode_2} = \aket{0}\abra{0}\otimes\aket{0}\abra{0} & &\Longleftrightarrow & &\map{\DCTCsCRMap}_{\UnitaryMultimode_2} = \aket{\qClock_2^{(0)}}\abra{\qClock_2^{(0)}}, \nonumber\\
	&\text{one clock:} & &\map{\DCTCsCVMap}_{\UnitaryMultimode_2} = \left\{\substack{ \aket{\qclock^{(1)}}\abra{\qclock^{(1)}}\otimes\aket{0}\abra{0} \\ \aket{0}\abra{0}\otimes\aket{\qclock^{(1)}}\abra{\qclock^{(1)}}}\right\} & &\Longleftrightarrow & &\map{\DCTCsCRMap}_{\UnitaryMultimode_2} = \aket{\qClock_2^{(1)}}\abra{\qClock_2^{(1)}}, \nonumber\\
	&\text{two clocks:} & &\map{\DCTCsCVMap}_{\UnitaryMultimode_2} = \aket{\qclock^{(1)}}\abra{\qclock^{(1)}}\otimes\aket{\qclock^{(2)}}\abra{\qclock^{(2)}} & &\Longleftrightarrow & &\map{\DCTCsCRMap}_{\UnitaryMultimode_2} = \aket{\qClock_2^{(2)}}\abra{\qClock_2^{(2)}},
\end{align}
\end{widetext}
whereby the weighting outside of each CR term in the full mixture is simply the sum of the $\Free_{u,v}$ coefficients which contribute to it. Similar to the CV case (\ref{eq:multimode_D-CTCs_CV_together}), this $2$-mode idea provides the recipe for state construction and can therefore be readily extended to the $\NumberModes$-mode CR state, yielding
\begin{align}
	\map{\DCTCsCRMap}_{\UnitaryMultimode_\NumberModes} = \sum_{\gv{\alpha} = \v{0}}^{\v{1}} \Free_{\gv{\alpha}} \aket{\qClock_\NumberModes^{(\abs{\gv{\alpha}})}}\abra{\qClock_\NumberModes^{(\abs{\gv{\alpha}})}}. \label{eq:multimode_D-CTCs_CR_together}
\end{align}
From this, it is easy to see that, while each term in the trapped state (\ref{eq:multimode_D-CTCs_CV_together}) is unique in structure (temporal ordering) of the clocks, they are degenerate in the order of their most evolved clock. In other words, multiple individual fixed points may contribute to the same term in the output state (\ref{eq:multimode_D-CTCs_CR_together}).

\subsubsection{Equivalent circuit picture}\label{sec:ECP}
Despite rectifying the classical indeterminism of the problem, the multiplicity in the solutions to the D-CTC CV state (\ref{eq:multimode_D-CTCs_CV_together}) establishes another kind of indeterminism into the mix. Deutsch's original correction for this issue was to argue that the ``correct'' solution is the one with the most entropy. Appealing to thermodynamics, he reasoned that the trapped CTC state would tend towards a maximally entropic equilibrium state. The lack of rigorous theoretical support to such a claim however led to the formulation of the equivalent circuit picture (ECP) \cite{ralph_information_2010,ralph_reply_2011,pienaar_quantum_2011,ralph_relativistic_2012,dong_ralphs_2017}, in which the form of the D-CTC trapped state becomes intuitive.

In the ECP, the CV state may be ascertained by iterating an initial ``seed'' state through the circuit \emph{ad infinitum}. Upon reaching the fixed point in the limit, the corresponding CR output state may be calculated. For our purposes, we will use the normalised, arbitrary classical mixture in the clock space
\begin{equation}
	\varrho \equiv \sum_{n=1}^{\NumberLevels} w_n\aket{n}\abra{n},
\end{equation}
to define the ``seed'' state of the CV system as
\begin{equation}
	\rho = \left[\Free\aket{0}\abra{0} + (1-\Free)\varrho\right]^{\otimes\NumberModes}. \label{eq:multimode_seed}
\end{equation}
As a consequence of the linearity of the circuit, the separate components in the end state fixed point (\ref{eq:multimode_D-CTCs_CV_together}) are linear transformations of the inputs components here in the seed state. This means that the mixture fixes the free parameters such that
\begin{equation}
	\Free_{\gv{\alpha}} = \Free^{\NumberModes - \abs{\gv{\alpha}}} (1-\Free)^{\abs{\gv{\alpha}}}. \label{eq:D-CTCs_ECP_free}
\end{equation}
The associated output state (\ref{eq:multimode_D-CTCs_CR_together}) is thus
\begin{align}
	\map{\DCTCsCRMap}_{\UnitaryMultimode_\NumberModes}^\mathrm{ECP} &= \sum_{\gv{\alpha}= \v{0}}^{\v{1}} \Free^{\NumberModes - \abs{\gv{\alpha}}} (1-\Free)^{\abs{\gv{\alpha}}}  \aket{\qClock_\NumberModes^{(\abs{\gv{\alpha}})}}\abra{\qClock_\NumberModes^{(\abs{\gv{\alpha}})}} \nonumber\\
	&= \sum_{k=0}^{\NumberModes} \binom{\NumberModes}{k}\Free^{\NumberModes - k} (1-\Free)^{k} \aket{\qClock_\NumberModes^{(k)}}\abra{\qClock_\NumberModes^{(k)}}. \label{eq:multimode_D-CTCs_CR_together_equivalent}
\end{align}
As an example, it is straightforward to analytically verify that the seed state (\ref{eq:multimode_seed}) in the $\NumberModes = 2$ case yields
\begin{subequations}
	\begin{align}
		\Free_{0,0} &= \Free^2, \\
		\Free_{1,0} = \Free_{0,1} &= \Free(1-\Free), \\
		\Free_{1,1} &= (1-\Free)^2.
	\end{align}
\end{subequations}

\subsection{P-CTC solutions}\label{sec:results_P-CTCs}

For P-CTCs, given the complicated form of the unitary (\ref{eq:unitary_multimode}), it is intractable to compute the general $\NumberModes$-mode output state by projecting on to the CV modes as per (\ref{eq:P-CTCs_map}). It is feasible however to construct the solution by quantifying the action of the unitary on an input clock. We can do this by first considering an input clock state $\aket{\qclock}$ in isolation on a single mode (say, the $m$th one) in the circuit, that is, we are interested how the state $\aket{0}^{\otimes(m-1)}\otimes\aket{\qclock}\otimes\aket{0}^{\otimes(\NumberModes-m)}$ evolves. Observing figure \ref{fig:circuit_multimode}, it is easy to see that such an input would only interact with the $m$th set of \textgate{swap} gates, and so we merely have to solve this reduced circuit, i.e., compute the action of the partial unitary
\begin{align}
	\op{\UnitaryTraced}_\NumberModes^{(m)} &= \tr_\CV\Bigl[\left(\op{\vac{\Identity}}^{\otimes\NumberModes} \otimes \op{\vac{\Rotation}}^{\otimes\NumberModes}(\DeltaTime)\right) \nonumber\\
	&\qquad\qquad \left(\op{\vac{\Swap}}_{m,2\NumberModes} \ldots \op{\vac{\Swap}}_{m,(\NumberModes+2)} \, \op{\vac{\Swap}}_{m,(\NumberModes+1)}\right)\Bigr] \label{eq:unitary_multimode_isolated}
\end{align}
on the clock state, as per the P-CTC prescription. The action of this reduced unitary can be evaluated by following the flow of the state through the circuit while noting a few important rules. Each time the clock meets a $\textgate{swap}$ gate whose paired CR mode is nonvacuous, the CV clock will swap with its CR counterpart. As the CV clock will be one ``tick'' ahead, this effect manifests as the contribution of a factor of $\op{\Rotation}$ in the partial trace operation. Alternatively, if the CR mode is vacuous, then the vacuum trace over the time evolution gate yields the coefficient $\abra{0}\op{\Rotation}\aket{0}$ [which is unity as per (\ref{eq:rotation_vacuum_unity})]. All together, the partial trace (\ref{eq:unitary_multimode_isolated}) for any single \textgate{swap} is then the sum of these contributions, i.e., $\op{\Identity} + \op{\Rotation}$, which of course simply gives the output state $\aket{\qclock^{(0)}} + \aket{\qclock^{(1)}}$. Since there are $\NumberModes$ such interactions in series on the $m$th mode, then the effect of the superposition of nonvacuous and vacuous interactions compounds such that one obtains the binomially distributed superposition of evolutions
\begin{equation}
	\op{\UnitaryTraced}_\NumberModes^{(m)}\aket{\qclock} = \left(\op{\Identity} + \op{\Rotation}\right)^\NumberModes\aket{\qclock} = \sum_{k=0}^{\NumberModes} \binom{\NumberModes}{k}\aket{\qclock^{(k)}}. \label{eq:unitary_multimode_isolated_single}
\end{equation}
Because a clock on the $m$th CR mode does not interact with any \textgate{swap} gate not connected to this mode, then the components of the total unitary (\ref{eq:unitary_multimode}) which act on other modes have no action on this state. As the input form (\ref{eq:clock_multimode}) is an entangled state between a clock and the vacuum, this means that the modes act in perfect isolation. Mathematically, each mode is effectively independent, and so the complete P-CTC output can be deduced to be
\begin{align}
	\map{\PCTCsMap}_{\UnitaryMultimode_\NumberModes}\MapBracket{\op{\CRState}}(\DeltaTime) &\sim \op{\UnitaryTraced}_\NumberModes\aket{\qClock_\NumberModes} \nonumber\\
	&= \sum_{k=0}^{\NumberModes} \binom{\NumberModes}{k} \aket{\qClock^{(k)}_\NumberModes} \label{eq:multimode_P-CTCs_output}.
\end{align}
Essentially, a circuit with $\NumberModes$ modes has $\binom{\NumberModes}{m}$ possible ways in which a clock may time travel $m$ times during its history, and so the corresponding component in the output is weighted accordingly.

\subsection{Probabilities}

It is often useful to calculate probabilities in order to gain a more intuitive understanding of what the D-CTC (\ref{eq:multimode_D-CTCs_CR_together}) and P-CTC (\ref{eq:multimode_P-CTCs_output}) output states physically represent in terms of chance and outcome. This is accomplished by computing specific clock overlaps. For D-CTCs in the case that the input clock's orthogonalisation time is matched to the wormhole's time delay, we can compute the probability of measuring a clock of $k$ evolutions in the output state to be
\begin{equation}
	\left.\abra{\qClock^{(k)}_\NumberModes(\OrthogonalisationTime)} \map{\DCTCsCRMap}_{\UnitaryMultimode_\NumberModes} \aket{\qClock^{(k)}_\NumberModes(\OrthogonalisationTime)}\right|_{\DeltaTime=\OrthogonalisationTime} = \sum_{\substack{\gv{\alpha} = \v{0} \\ \abs{\gv{\alpha}} = k}}^{\v{1}} \Free_{\gv{\alpha}} \label{eq:D-CTCs_probabilities}
\end{equation}
which, as expected, merely represents the count of clocks which have evolved $k$ times through the wormhole. In the case of the ECP (Sec.~\ref{sec:ECP}), the fixing of the free parameters $\Free_{\gv{\alpha}}$ as per (\ref{eq:D-CTCs_ECP_free}) means that (\ref{eq:D-CTCs_probabilities}) reduces to
\begin{equation}
	\left.\abra{\qClock^{(k)}_\NumberModes(\OrthogonalisationTime)} \map{\DCTCsCRMap}_{\UnitaryMultimode_\NumberModes} \aket{\qClock^{(k)}_\NumberModes(\OrthogonalisationTime)}\right|_{\DeltaTime=\OrthogonalisationTime} = \binom{\NumberModes}{k}\Free^{\NumberModes - k} (1-\Free)^{k}. \label{eq:D-CTCs_probabilities_ECP}
\end{equation}
For P-CTCs, we can similarly calculate the $k$ evolutions probabilities as
\begin{equation}
	\left.\abra{\qClock^{(k)}_\NumberModes(\OrthogonalisationTime)} \map{\PCTCsMap}_{\UnitaryMultimode_\NumberModes} \aket{\qClock^{(k)}_\NumberModes(\OrthogonalisationTime)}\right|_{\DeltaTime=\OrthogonalisationTime} \propto {\binom{\NumberModes}{k}}^2. \label{eq:P-CTCs_probabilities}
\end{equation}
Perhaps the most interesting observation of these two sets of probabilities is their stark physical difference. While both do contain the same binomial coefficient (interpretable as a factor of multiplicity in the loop number), the D-CTC set (\ref{eq:D-CTCs_probabilities_ECP}) is a binomial distribution dependent on the free parameter $\Free$, while the P-CTC probabilities (\ref{eq:P-CTCs_probabilities}) are simply constant for any fixed $k$ and $\NumberModes$.

\section{Behaviour in the limit of large $\NumberModes$}\label{sec:limits}

The regime explored until now concerns a positive, finite, integer $\NumberModes$ number of CR modes. Physically, this corresponds to a situation in which the particle can travel along a finite number of trajectories. While interesting, a more realistic case would be the continuum limit, namely, to allow the particle to occupy any position in space (as constrained by the one-dimensional model). Such a situation necessitates a continuum of modes, which coincides with the limiting behaviour of the model, i.e., when $\NumberModes$ goes to infinity.

\subsection{Limiting behaviour for D-CTCs}

For D-CTCs, for simplicity we shall consider only the case where all free parameters (i.e., the $\Free$ coefficients) take the same value (as per the ECP result in Sec.~\ref{sec:ECP}). From (\ref{eq:multimode_D-CTCs_CR_together_equivalent}), the probability for a clock of $k$ ticks to be measured in the output state is
\begin{equation}
	\mathrm{Pr}_\mathrm{D}(k;\NumberModes,\Free) = \binom{\NumberModes}{k}\Free^{\NumberModes - k} (1-\Free)^{k}. \label{eq:probability_D-CTCs_ECP}
\end{equation}
It is easy to then verify that this diverges as $\NumberModes$ becomes very large, which means that the average number of clocks inhabiting the CTC likewise blows up. This is of course problematic, as it suggests that a large number of modes necessarily corresponds to an unbounded quantity of energy (in the form of clocks) inside the wormhole.

One would rather expect that the average clock number to remain finite. To accomplish this, we set
\begin{equation}
	\Free = \FreeScaled^{1/\NumberModes} \label{eq:free_scaled}
\end{equation}
for some $0 \leq \FreeScaled \leq 1$. This implicitly defines $\FreeScaled = \Free^\NumberModes$ which, according to the ECP parameters as per (\ref{eq:D-CTCs_ECP_free}), corresponds to the probability that all CV modes are vacuous. Note that if $\FreeScaled=0$ then of course $\Free = 0$ for all $\NumberModes$. This is not a particularly meaningful case, as it corresponds to all CV modes being occupied and therefore the probabilities (\ref{eq:probability_D-CTCs_ECP}) still diverge with $\NumberModes$. Alternatively, when $\FreeScaled \neq 0$, for which there is an infinite number of possible values $0 < \FreeScaled \leq 1$, one has $\Free \rightarrow 1$ as $\NumberModes \rightarrow \infty$. This suggests that higher order clocks become less probable as $\NumberModes$ grows because the likelihood of any CV mode being vacuous increases. Physically, we argue that this corresponds to energy and/or space constraints in the wormhole --- finite-size billiard balls occupy volume in space, and so there must be a limit to the number of balls that can simultaneously exist inside. The dimensionless parameter $\FreeScaled$ can thus be thought of as a regulator for energy inside the wormhole.

\begin{figure}
	\includegraphics[scale=1.15]{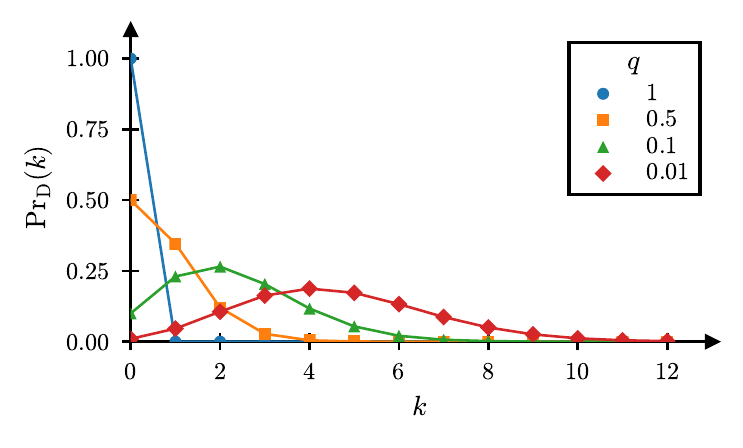}
	\caption[Distribution of clock evolution probabilities for D-CTCs]{\label{fig:distribution_D-CTCs}Plot of the probability Eq.~(\ref{eq:probability_D-CTCs_limit}) that a clock measured in the D-CTC output evolved $k$ times through the time machine. Here, $\FreeScaled$ is the probability that, in the seed state, all CV modes are vacuous.}
\end{figure}

In any case, we now note that, with (\ref{eq:free_scaled}), the probability (\ref{eq:probability_D-CTCs_ECP}) becomes
\begin{equation}
	\mathrm{Pr}_\mathrm{D}(k;\NumberModes,\FreeScaled) = \binom{\NumberModes}{k}\FreeScaled^{1 - k/\NumberModes} (1-\FreeScaled^{1/\NumberModes})^{k}
\end{equation}
which converges in the large $\NumberModes$ limit to the value
\begin{equation}
	\mathrm{Pr}_\mathrm{D}(k;\FreeScaled) = \lim\limits_{\NumberModes\rightarrow\infty}\mathrm{Pr}_\mathrm{D}(k;\NumberModes,\FreeScaled) = \frac{\FreeScaled \left(\ln \frac{1}{\FreeScaled}\right)^k}{k!}. \label{eq:probability_D-CTCs_limit}
\end{equation}
Making the change of variables
\begin{equation}
	\lambda = \ln \frac{1}{\FreeScaled} \Leftrightarrow \FreeScaled = \e^{-\lambda}
\end{equation}
subsequently reveals that the clock tick probabilities have a Poisson distribution, i.e.,
\begin{equation}
	\mathrm{Pr}_\mathrm{D}(k;\lambda) = \frac{\lambda^{k}\e^{-\lambda}}{k!} = \text{Pois}(k;\lambda).
\end{equation}
This tells us that, as we desired, the expectation value for the clock's tick count is no longer infinite, which is confirmed by the result
\begin{equation}
	\mathrm{E}_\mathrm{D}\left[k\right] = \lim\limits_{\NumberModes\rightarrow\infty} \sum_{k=0}^{\NumberModes} k \, \mathrm{Pr}_\mathrm{D}(k;\NumberModes,\FreeScaled) = \ln \frac{1}{\FreeScaled} = \lambda.
\end{equation}
Figure \ref{fig:distribution_D-CTCs} visualises the discrete probability distribution of Eq.~(\ref{eq:probability_D-CTCs_limit}) in the continuum limit for example values of $\FreeScaled$.

\subsection{Limiting behaviour for P-CTCs}

For P-CTCs, the probabilities from (\ref{eq:multimode_P-CTCs_output}) are
\begin{equation}
	\mathrm{Pr}_\mathrm{P}(k;\NumberModes) = \frac{\binom{\NumberModes}{k}^2}{\sum_{m=0}^{\NumberModes}\binom{\NumberModes}{m}^2} = \frac{\binom{\NumberModes}{k}^2}{\binom{2\NumberModes}{\NumberModes}}, \label{eq:probability_P-CTCs}
\end{equation}
and so the expectation of measuring $k$-evolution clocks takes the intuitive value
\begin{equation}
	\mathrm{E}_\mathrm{P}\left[k;\NumberModes\right] = \sum_{k=0}^{\NumberModes} k \, \mathrm{Pr}_\mathrm{P}(k;\NumberModes) = \frac{\NumberModes}{2}.
\end{equation}
Given that this value is expected to remain finite, the fact that it diverges as $\NumberModes \rightarrow \infty$ calls into question the validity of taking such a limit in the context of our model. There are two main conclusions one can therefore reach from this result: either the circuit model itself is pathological, or P-CTCs require modification in order to mitigate the continuum issue. For the former, the expectation that a sensible, well-defined continuum limit should be obtainable may indeed be much too na\"{i}ve. Given that D-CTCs required regularisation, then it is not unreasonable to anticipate that the postselected teleportation prescription may likewise require modification to either moderate the result here (thereby providing convergence), or to truncate the output state (\ref{eq:multimode_P-CTCs_output}). Here, we propose two ways by which one can guarantee convergence in the continuum limit.

The first is to introduce a regularisable parameter into the P-CTC approach similar to the $\Free$ parameter for D-CTCs. This can be achieved if we relax the requirement in the P-CTC prescription which demands both the postselected and prepared states to be maximally entangled. Ordinarily, one uses the product state $\op{\Identity}^{\otimes\NumberModes}\otimes\aket{\Phi^+}$ where $\aket{\Phi^+}$ is a Bell state, as this corresponds to complete teleportation. However, self-consistent output states can still be procured when the entanglement is not necessarily maximal but takes the general form
\begin{equation}
	\aket{\Phi^+(\Incomplete)} = \left(\sqrt{\Incomplete}\aket{0}\otimes\aket{0} + \sqrt{1-\Incomplete} \frac{1}{\sqrt{\NumberLevels}}\sum_{n=1}^{\NumberLevels}\aket{n}\otimes\aket{n}\right)^{\otimes\NumberModes} \label{eq:Bell_incomplete}
\end{equation}
where $0\leq \Incomplete \leq 1$ is an independent parameter. Deviations of $\Incomplete$ away from $\frac{1}{\NumberLevels+1}$ (the value of maximal entanglement) correspond to incomplete teleportation, wherein there is a probability that the state does not interact with the CTC. Note that this form (\ref{eq:Bell_incomplete}) does not converge to a valid state as $\NumberModes \rightarrow \infty$ (in the sense that there is no Hilbert-space vector to which the sequence converges). We therefore argue that the model need only be defined relative to a physical state, meaning that only the outcomes of measurements are required to remain meaningful (i.e., converge) in the continuum limit.

After many interactions, the effect of incomplete teleportation compounds, resulting in the suppression of highly evolved clocks. Mathematically, one obtains the output state
\begin{equation}
	\map{\PCTCsMap}_{\UnitaryMultimode_\NumberModes}\MapBracket{\op{\CRState}}(\DeltaTime) \sim \sum_{k=0}^{\NumberModes} \binom{\NumberModes}{k} \Incomplete^{\NumberModes-k}\frac{(1-\Incomplete)^{k}}{\NumberLevels^k} \aket{\qClock^{(k)}_\NumberModes} \label{eq:incomplete_teleportation}
\end{equation}
from which it is easy to see that $\Incomplete$ acts much like the free parameter $\Free$ from D-CTCs (\ref{eq:multimode_D-CTCs_CR_together_equivalent}). Indeed, given the normalised probabilities
\begin{equation}
	\mathrm{Pr}_{\mathrm{P},\Incomplete}(k;\NumberModes,\Incomplete) = \frac{\left[\frac{\Incomplete^{\NumberModes}}{\NumberLevels^{k}}\binom{\NumberModes}{k}\left(\frac{1-\Incomplete}{\Incomplete}\right)^k\right]^2}{\sum_{m=0}^{\NumberModes}\left[\frac{\Incomplete^{\NumberModes}}{\NumberLevels^{m}}\binom{\NumberModes}{m}\left(\frac{1-\Incomplete}{\Incomplete}\right)^m\right]^2}, \label{eq:probability_P-CTCs_incomplete}
\end{equation}
it is straightforward to check that the corresponding expectation value does not remain finite in the continuum limit for general $\Incomplete$ and $\NumberLevels$. However, if the number of clock energy levels (equal to the number of orthogonal clock states) scales with the number of modes, i.e., $\NumberLevels = \NumberModes$, then we do in fact procure convergence in the continuum limit. In other words, when we grant the clock the ability to distinguish between every distinct path in the circuit, P-CTCs become regulated for any valid $\Incomplete$. As such, we can calculate the associated normalised probability distribution in the limit $\NumberModes\rightarrow\infty$ to be
\begin{equation}
	\mathrm{Pr}_{\mathrm{P},\Incomplete}(k;\Incomplete) = \frac{1}{I_0\left(2\frac{1 - \Incomplete}{\Incomplete}\right)}\frac{1}{(k!)^2}\left(\frac{1 - \Incomplete}{\Incomplete}\right)^{2k}, \label{eq:probability_P-CTCs_incomplete_limit}
\end{equation}
with the expectation
\begin{equation}
	\mathrm{E}_{\mathrm{P},\Incomplete}\left[k;\Incomplete\right] = \frac{1 - \Incomplete}{\Incomplete}\frac{I_1\left(2\frac{1 - \Incomplete}{\Incomplete}\right)}{I_0\left(2\frac{1 - \Incomplete}{\Incomplete}\right)}. \label{eq:expectation_P-CTCs_incomplete_limit}
\end{equation}
Here, $I_n$ is the modified Bessel function of order $n$. From this, we can conclude that the introduction of incomplete teleportation via the parameter $\Incomplete$ in the entangled state, combined with the requirement that the clock states scale with the modes, are together sufficient to guarantee well-behaved probabilities (see figure 6).

\begin{figure}
	\includegraphics[scale=1.15]{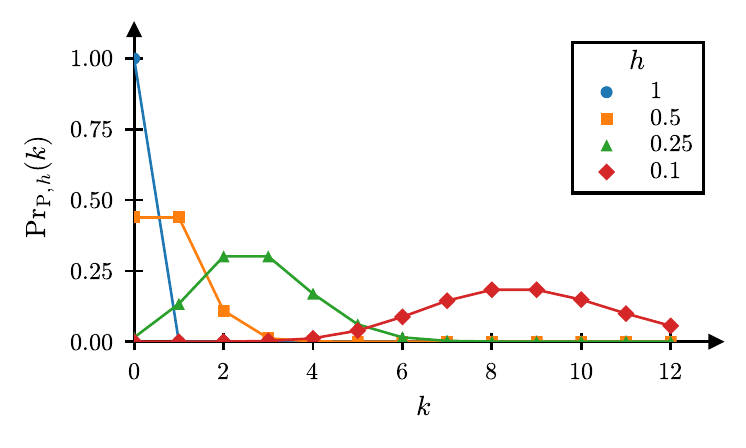}
	\caption[Distribution of clock evolution probabilities for P-CTCs_incomplete]{\label{fig:distribution_P-CTCs_incomplete}Plot of the probability Eq.~(\ref{eq:probability_P-CTCs_incomplete_limit}) that a clock measured in the P-CTC output evolved $k$ times through the time machine. Here, $\Incomplete$ is a parameter that regularises the postselected state, as shown in Eq.~(\ref{eq:Bell_incomplete}).}
\end{figure}

\begin{figure}[h]
	\hspace*{0cm}\includegraphics{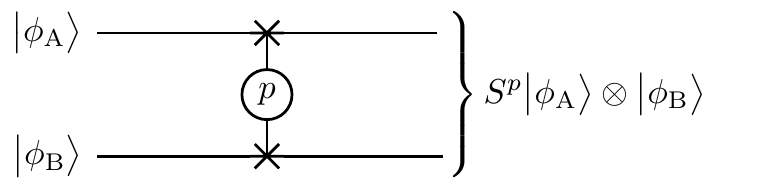}
	\caption[Dispersion]{\label{fig:dispersion_gate}Pictorial representation of the power of \textgate{swap} gate.}
\end{figure}

The second way to provide convergence in the continuum limit is to supplant the deterministic vacuum-excluding \textgate{swap} gate with a probabilistic variant. This is motivated by particle scattering theory, in which interactions have some finite probability for the particles to miss each other. Since our hard-sphere collision here is an approximation to some underlying, fundamental scattering event, then in general we expect there to be a finite probability for which it does not occur. In perhaps the simplest model, the spontaneous, probabilistic diffusion of a bipartite quantum state between a pair of modes may be characterised by a probabilistic \textgate{swap} \cite{williams_explorations_2010},
\begin{equation}
	\op{\Swap}^\Power = \frac{1 + \e^{-\eye\pi \Power}}{2}\op{\Identity}\otimes\op{\Identity} + \frac{1 - \e^{-\eye\pi \Power}}{2}\op{\Swap}. \label{eq:SWAP_power}
\end{equation}
In this form, the unitary is simply the ordinary \textgate{swap} gate $\op{\Swap}$ of Eq.~(\ref{eq:SWAP}) taken to the power $\Power \in \mathds{R}$. Here, the parameter $\Power$ represents the extent to which the \textgate{swap} will occur: complete exchange will only be produced when $\Power$ is an odd integer, no interaction occurs for even integers, and non-integer values correspond to partial collisions. In essence, this allows us to vary the ``strength'' of the interaction. For example, given an input state $\aket{\qclock_\textrm{A}} \otimes \aket{\qclock_\textrm{B}}$, the power \textgate{swap} gate has the action
\begin{align}
	\op{\Swap}^\Power \aket{\qclock_\textrm{A}} \otimes \aket{\qclock_\textrm{B}} &= \frac{1 + \e^{-\eye\pi \Power}}{2} \aket{\qclock_\textrm{A}} \otimes \aket{\qclock_\textrm{B}} \nonumber\\
	&\quad + \frac{1 - \e^{-\eye\pi \Power}}{2} \aket{\qclock_\textrm{B}} \otimes \aket{\qclock_\textrm{A}},
\end{align}
and so the power $\Power$ may be interpreted precisely as the probability of intermodal exchange. This may be represented diagrammatically as per figure \ref{fig:dispersion_gate}.

For our purposes, it is convenient to define the functions
\begin{subequations}
\begin{align}
	\ProbabilisticFalse &= \frac{1 + \e^{-\eye\pi \Power}}{2},\\
	\ProbabilisticTrue &= \frac{1 - \e^{-\eye\pi \Power}}{2},
\end{align}
\end{subequations}
so that $\abs{\ProbabilisticFalse} + \abs{\ProbabilisticTrue} = 1$. With these, we may for example write the $\NumberModes$-mode model's probabilistic \textgate{swap} interaction between modes $i,j$ as
\begin{equation}
	\op{\vac{\Swap}}^\Power_{i,j} = \ProbabilisticFalse\op{\vac{\Identity}}^{\otimes\NumberModes} + \ProbabilisticTrue\op{\vac{\Swap}}_{i,j} \label{eq:SWAP_probabilistic}
\end{equation}
so that $\ProbabilisticTrue$ is the probability amplitude which corresponds positively to interaction. When we replace every gate in the circuit's unitary (\ref{eq:unitary_multimode}) with such a form, we obtain exactly $2^{M^2}$ terms. For large $\NumberModes$, the complexity of the circuit prevents a rigorous approach to determining the P-CTC solution, but fortunately we can use a few cunning techniques in order expedite the process of determining a nonrigorous alternative. First, consider a single term in the input state (\ref{eq:clock_multimode}), e.g., $\aket{0}^{\otimes(n-1)}\otimes\aket{\qclock}\otimes\aket{0}^{\otimes(\NumberModes-n)}$, which represents the particle being located in the $n$th mode. For such a state, it is easy to conclude that the particle would only interact with \textgate{swap} gates which act on the $n$th mode, and so any other gates leave the state unaffected. Accordingly, when one expands the circuit's unitary into its $2^{M^2}$ terms, only those which possess gates of the form $\op{\vac{\Swap}}_{n,m}$ and/or $\op{\vac{\Swap}}_{m,n}$ need to be considered for the $n$th mode term. With this in mind, the circuit is simple enough to solve for small $\NumberModes$ using the postselected teleportation interpretation of P-CTCs. In doing so, we are able to formulate a few common rules:
\begin{enumerate}[label=(\roman*),itemsep=0.015cm]
	\item The count of a clock's time evolutions corresponds to the number of loops it takes through the CTC.
	\item The number of loops which a clock takes through the CTC corresponds to the number of \textgate{swap} gates with which it interacts with.
	\item Every instance in which the clock reaches a \textgate{swap} gate whose paired mode is nonvacuous, it has the probability amplitudes $\ProbabilisticTrue$ and $\ProbabilisticFalse$ to swap and not swap respectively, and picks up such factors in each case.
	\item When the clock reaches a \textgate{swap} gate and is not knocked into the CTC (due to the paired mode being in the vacuum), it picks up a factor of $\abra{0}\op{\vac{\Rotation}}\aket{0}$.
	\item Every loop which the clock cannot interact with contributes a factor of $\tr[\op{\vac{\Rotation}}]$ to the clock.
\end{enumerate}
Essentially, the first condition on a successful interaction at a \textgate{swap} gate is that the paired mode must be occupied. If it is, the particle then has a $\ProbabilisticTrue$ probability amplitude to swap with the other particle (and an $\ProbabilisticFalse$ probability amplitude to not swap). Otherwise, the paired mode must be vacuous, and accordingly the clock passes through unperturbed and picks up a vacuous factor $\abra{0}\op{\vac{\Rotation}}\aket{0}$. Any CV mode which the particle cannot interact with contributes a factor of $\tr[\op{\vac{\Rotation}}]$ to the relevant term, as a closed loop containing any (single-mode) gate corresponds to the full trace of that gate. Using such rules, one can construct the output state which corresponds to any single-particle input state.

\begin{figure}
	\includegraphics[scale=1.15]{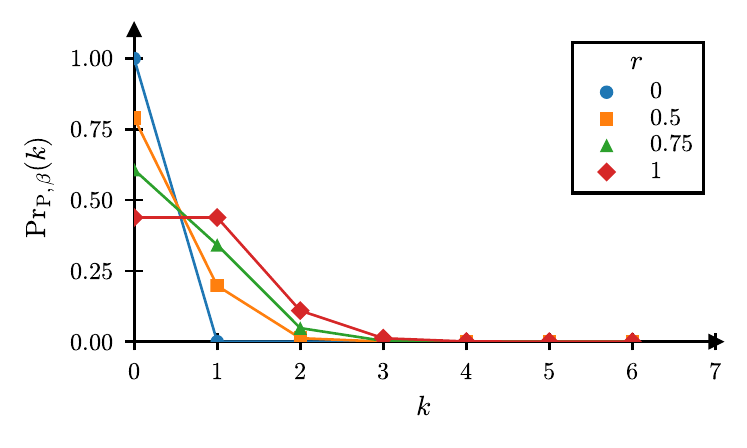}
	\caption[Distribution of clock evolution probabilities for P-CTCs_power]{\label{fig:distribution_P-CTCs_power}Plot of the probability Eq.~(\ref{eq:probability_P-CTCs_power_limit}) that a clock measured in the P-CTC output evolved $k$ times through the time machine. Here, the P-CTC is regularised by first replacing the hard-sphere interactions with partial collisions parametrised by $\ProbabilisticTrue$ [see Eq.~(\ref{eq:SWAP_probabilistic})], followed by introducing the rescaled parameter $\ProbabilisticTrueScaled$ as per Eq.~(\ref{eq:regularisation_power}).}
\end{figure}

By putting everything together for all other particle states of the input entangled state, we conjecture the $\NumberModes$-mode P-CTC output to be
\begin{align}
	\map{\PCTCsMap}_{\UnitaryMultimode_\NumberModes^p}\MapBracket{\op{\CRState}}(\DeltaTime) \sim \sum_{k=0}^{\NumberModes} \binom{\NumberModes}{k} &\left(\ProbabilisticFalse\left\{1 + \tr\left[\op{\Rotation}\right]\right\} + \ProbabilisticTrue\right)^{\NumberModes - k} \nonumber\\
	&\qquad\quad\:\:\times\ProbabilisticTrue^k \aket{\qClock^{(k)}_\NumberModes(\DeltaTime)}.
\end{align}
For orthogonal clocks, we have $\abraket{\qclock^{(k)}(\OrthogonalisationTime)}{\qclock^{(k+1)}(\OrthogonalisationTime)} = \tr\left[\op{\Rotation}\right] = 0$, and one thus obtains
\begin{equation}
	\map{\PCTCsMap}_{\UnitaryMultimode_\NumberModes^p}\MapBracket{\op{\CRState}}(\OrthogonalisationTime) \sim \sum_{k=0}^{\NumberModes} \binom{\NumberModes}{k} \ProbabilisticTrue^k \aket{\qClock^{(k)}_\NumberModes(\OrthogonalisationTime)}.
\end{equation}
The (unnormalised) probability of measuring a clock of $k$ evolutions in the output state is $\left[\binom{\NumberModes}{k} \ProbabilisticTrue^{k}\right]^2$, and so the normalised probability distribution is
\begin{equation}
	\mathrm{Pr}_{\mathrm{P},\ProbabilisticTrue}(k;\NumberModes,\ProbabilisticTrue) = \frac{\left[\binom{\NumberModes}{k} \ProbabilisticTrue^{k}\right]^2}{\sum_{m=0}^{\NumberModes}\left[\binom{\NumberModes}{m} \ProbabilisticTrue^{m}\right]^2}. \label{eq:probability_P-CTCs_power}
\end{equation}
It is easy to verify that this does not correspond to a finite expectation value for general $\ProbabilisticTrue$, and so, like previously, we must introduce regularisation to make sense in the continuum limit. As the binomial coefficient is bound from above by
\begin{equation}
	\binom{\NumberModes}{k} \leq \frac{\NumberModes^k}{k!},
\end{equation}
we can set
\begin{equation}
	\ProbabilisticTrue = \frac{\ProbabilisticTrueScaled}{\NumberModes} \label{eq:regularisation_power}
\end{equation}
in order to guarantee convergence by eliminating all instances of $\NumberModes$ in the upper bound of the product $\binom{\NumberModes}{k}\ProbabilisticTrue^k$. It is then straightforward to deduce that the normalised probability in the $\NumberModes \rightarrow \infty$ limit takes the form
\begin{equation}
	\mathrm{Pr}_{\mathrm{P},\ProbabilisticTrue}(k;\ProbabilisticTrueScaled) = \frac{1}{I_0(2\ProbabilisticTrueScaled)}\frac{\ProbabilisticTrueScaled^{2k}}{(k!)^2}. \label{eq:probability_P-CTCs_power_limit}
\end{equation}
The corresponding expectation value is therefore
\begin{equation}
	\mathrm{E}_{\mathrm{P},\ProbabilisticTrue}\left[k;\ProbabilisticTrueScaled\right] = \ProbabilisticTrueScaled\frac{I_1(2\ProbabilisticTrueScaled)}{I_0(2\ProbabilisticTrueScaled)}. \label{eq:expectation_P-CTCs_power_limit}
\end{equation}
A plot of the probability distribution (\ref{eq:probability_P-CTCs_power_limit}) appears in figure \ref{fig:distribution_P-CTCs_power}. We conclude that, in the scenario for which the interactions are probabilistic, the evolution information extracted from the clock can be made meaningful in the continuum limit only when the chance of self-collision is made to inversely scale with the number of modes.

\section{Other Issues}\label{sec:other}

\subsection{Distinction of the CTCs}

An important point which we reiterate here is that the physical scenario that our model corresponds to is one in which the individual CTCs, separated either spatially or temporally, all belong to the same wormhole. This is to say that both the clock entanglement and wormhole, spanning the numerous Hilbert spaces in the region, are not both so delocalised such that the CTCs may be taken as separate and ontologically distinct objects. Rather, while the Hilbert space of the CTCs is subdivisible, the entire collection must be treated as a single ``bundle'' of the lone wormhole.

If one however does treat the CTC's CV modes separately, then the circuit formulation functions essentially as $\NumberModes$ distinct CTCs. Since there are no interactions between the modes within either the CR or CV bundles, the associated D-CTC calculation is exceedingly straightforward, and one finds that there exist $\NumberModes$ CV states, with the $m$th one being directly dependent on the $(m-1)$th one. Let us first define the recursive state
\begin{equation}
	\map{\DCTCsMixing}_{m} = \Free_{m}\map{\DCTCsMixing}_{m-1} + (1 - \Free_{m})\op{\vac{\Rotation}}\map{\DCTCsMixing}_{m-1}\op{\vac{\Rotation}}^\dagger \label{eq:multimode_D-CTCs_single-mode_separate}
\end{equation}
where the initial value takes the form
\begin{equation}
	\map{\DCTCsMixing}_{1} = \Free_{1}\aket{\qclock}\abra{\qclock} + (1 - \Free_{1})\op{\vac{\Rotation}}\aket{\qclock}\abra{\qclock}\op{\vac{\Rotation}}^\dagger.
\end{equation}
Here, the free variables $\left\{\Free_m\right\}_{m=1}^{\NumberModes}$ arise as unrestricted parameters which specify the degree of mixing between the nonevolved and evolved states. Using this, we can track how the entangled state (\ref{eq:clock_multimode}) evolves through the circuit when we trace out each CV mode individually. The first CV state is simply
\begin{equation}
	\map{\DCTCsCVMap}_{1} = \Free_{1}\aket{0}\abra{0} + (1 - \Free_{1})\op{\vac{\Rotation}}\aket{\qclock}\abra{\qclock}\op{\vac{\Rotation}}^\dagger
\end{equation}
which seeds the general recurrence expression
\begin{equation}
	\map{\DCTCsCVMap}_{m} = \Free_{m}\aket{0}\abra{0} + (1 - \Free_{m})\op{\vac{\Rotation}}\map{\DCTCsMixing}_{m-1}\op{\vac{\Rotation}}^\dagger. \label{eq:multimode_D-CTCs_CV_separate}
\end{equation}
This is the trapped state that exists in the $m$th CV mode, and so (due to the structure of the model) it can contain clocks that have time evolved at most $m$ times through the CTC. These clocks are contained in the single-mode clock mixture (\ref{eq:multimode_D-CTCs_single-mode_separate}).

Likewise, the associated CR state after $m$ traces can be written recursively,
\begin{equation}
	\map{\DCTCsCRMap}_{m} = \Free_{m}\map{\DCTCsCRMap}_{m-1} + (1 - \Free_{m})\op{\vac{\Rotation}}^{\otimes\NumberModes}\map{\DCTCsCRMap}_{m-1}\op{\vac{\Rotation}}^{\dagger \otimes\NumberModes} \label{eq:multimode_D-CTCs_CR_separate}
\end{equation}
where
\begin{equation}
	\map{\DCTCsCRMap}_{1} = \Free_{1}\aket{\qClock_\NumberModes}\abra{\qClock_\NumberModes} + (1 - \Free_{1})\op{\vac{\Rotation}}^{\otimes\NumberModes}\aket{\qClock_\NumberModes}\abra{\qClock_\NumberModes}\op{\vac{\Rotation}}^{\dagger \otimes\NumberModes}
\end{equation}
is the initial value. It is easy to think of the recursive state definitions $\map{\DCTCsMixing}_m$ (\ref{eq:multimode_D-CTCs_single-mode_separate}) and $\map{\DCTCsCRMap}_m$ (\ref{eq:multimode_D-CTCs_CR_separate}) as expressing the same mixing effect, with the former being for the single-mode clock $\aket{\qclock}$ and the latter being for the multimode entangled clock $\aket{\qClock_\NumberModes}$. If we unravel out the recurrence for the final output state, we find
\begin{widetext}
\begin{align}
	\map{\DCTCsCRMap}_{\NumberModes} &= \Free_1\Free_2\ldots\Free_\NumberModes \aket{\qClock_\NumberModes}\abra{\qClock_\NumberModes} \nonumber\\
	&\quad + \left[(1-\Free_1)\Free_2\ldots\Free_\NumberModes + \Free_1(1-\Free_2)\ldots\Free_\NumberModes + \ldots + \Free_1\Free_2\ldots(1-\Free_\NumberModes) \right] \op{\vac{\Rotation}}^{\otimes\NumberModes}\aket{\qClock_\NumberModes}\abra{\qClock_\NumberModes}\op{\vac{\Rotation}}^{\dagger \otimes\NumberModes} \nonumber\\
	&\quad + \left[(1-\Free_1)(1-\Free_2)\ldots\Free_\NumberModes + \ldots \right] \left(\op{\vac{\Rotation}}^{\otimes\NumberModes}\right)^2\aket{\qClock_\NumberModes}\abra{\qClock_\NumberModes}\left(\op{\vac{\Rotation}}^{\dagger \otimes\NumberModes}\right)^2 \nonumber\\
	&\quad + \ldots \nonumber\\
	&\quad + (1-\Free_1)(1-\Free_2)\ldots(1-\Free_\NumberModes) \left(\op{\vac{\Rotation}}^{\otimes\NumberModes}\right)^\NumberModes\aket{\qClock_\NumberModes}\abra{\qClock_\NumberModes}\left(\op{\vac{\Rotation}}^{\dagger \otimes\NumberModes}\right)^\NumberModes
\end{align}
\end{widetext}
It is interesting to note that, according to the ECP (for which all the $\Free_m$ coefficients assume a uniform value $\Free$), the above output state is identical to its counterpart (\ref{eq:multimode_D-CTCs_CR_together_equivalent}).

For P-CTCs, treating the CTC modes as each belonging to their own wormhole has no influence on the output. This is easily proven, as the state $\aket{\qClock^{(k)}(\DeltaTime)}$, interacting with a single wormhole, evolves to become $\aket{\qClock^{(k)}(\DeltaTime)} + \aket{\qClock^{(k+1)}(\DeltaTime)}$. Thus, given the temporal ordering of the CTCs, the initial entanglement $\aket{\qClock}$ ``telescopes'' out via $\NumberModes$ wormholes and becomes the same state as the single wormhole case (\ref{eq:multimode_P-CTCs_output}).

\subsection{Dispersion}\label{sec:dispersion}
\begin{figure}
	\includegraphics[scale=1.35]{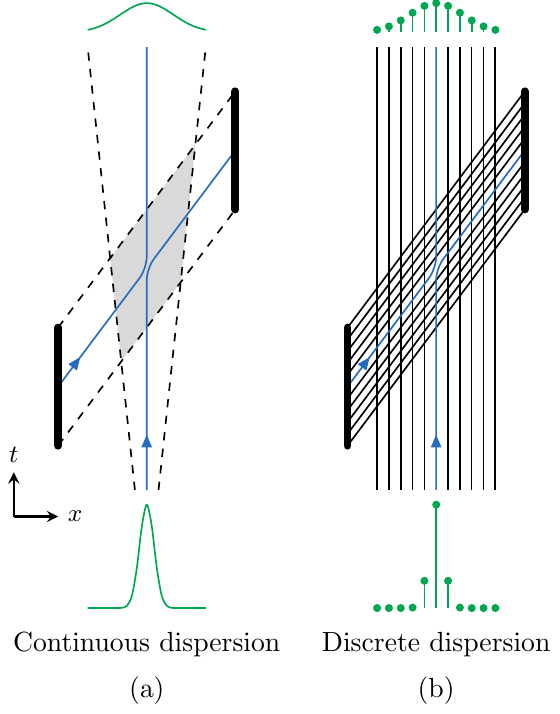}
	\caption[Dispersion]{\label{fig:dispersion_collection}Dispersion of a wave packet through a time-machine region. The wormhole's past and future mouths are the thick, solid lines on the left- and right-hand sides of each diagram, respectively.}
\end{figure}

A significant characteristic of quantum phenomena is dispersion in the wave packet. For simplicity, our base model described in Sec.~\ref{sec:model} does not capture dispersion, but we show here that it turns out to be inconsequential to the evolution of the particle. This is because, in the simplest case, a time-travelling quantum wave packet that disperses both in time and space must necessarily match with its past self (due to the self-consistency principle), thereby negating any additional dispersion acquired from evolution on the CTC. Figure \ref{fig:dispersion_collection} illustrates this concept in both the continuous and discrete pictures. Indeed, the absence of dispersion that originates from the CTC in the final output state can be mathematically shown to be a characteristic observed by both D-CTCs and P-CTCs.

For our circuit, dispersion may be introduced in the form of including probabilistic \textgate{swap} gates (\ref{eq:SWAP_power}) between neighbouring modes in the CR and CV bundles separately. With the addition of such a feature in the two-mode model (figure \ref{fig:circuit_two-mode}), it is a straightforward calculation to analytically verify that the output states remain unchanged compared to the nondispersive model for both D-CTCs and P-CTCs in the $\NumberModes=2$ and $\NumberModes=3$ cases. As such, it is reasonable to conclude that the same holds for a general number of modes. This supports the intuitive notion discussed above, wherein dispersion is negated simply by the requirement of self-consistency. However, as the gate described here is only relevant to (neighbouring) pairs of modes, it is possible that more general $\NumberModes$-mode dispersion does not completely vanish. Nevertheless, we speculate that this is not the case, though without such a unitary (the construction of which appears to be nontrivial), this is not possible to conclude with absolute certainty.

\section{Conclusion}\label{sec:conclusion}

In this paper, we introduced a simple formulation of an entirely quantum billiard-ball paradox. In contrast to most (if not all) past studies of this type, our model emulated the characteristically quantum evolution of a time-travelling wave packet by incorporating uncertainty into the localisation of the associated particle. As part of our model, we attached an internal degree of freedom to the particle. Acting like a clock, this incorporation allowed us to make measurements of the proper time of the distinct evolutions, and thereby distinguish between them. This enabled us to make statistical conclusions regarding the probability of each history. For example, we found that the D-CTC prescription robustly replicates the solution multiplicity of the billiard-ball paradox in the form of a set of mixed states. This solution, initially parametrised by $2^\NumberModes - 1$ free, arbitrary parameters, was demonstrated to be reducible to a one-parameter mixture in the equivalent circuit picture (ECP, see Sec.~\ref{sec:ECP}). P-CTCs on the other hand anticipate that all histories are equally likely, a result which agrees with both the conjecture of Friedman \emph{et al.} \cite{friedman_cauchy_1990} and the results of our earlier work \cite{bishop_time-traveling_2021} (regarding classical evolutions in the paradox).

Additionally, our results demonstrate that the solutions of both D-CTCs and P-CTCs each possess a distinct form of binomial distribution: as there are $\NumberModes$ modes both inside and outside the wormhole, then there are exactly $\binom{\NumberModes}{k}$ ways for $k$ interactions (and therefore loops) to occur. For D-CTCs, particularly in the ECP, this manifests plainly in the coefficients of each term in the solution (\ref{eq:multimode_D-CTCs_CR_together_equivalent}). Similarly, the weights attached to the distinct clock states in the P-CTC solution take on the form of binomial coefficients.

When we push the model further by exploring a continuum limit wherein there exists an infinite number of modes in the wormhole, we come to some rather significant conclusions. Through rescaling of the ECP free parameter, we showed how the D-CTC evolution probability distribution may be regularised so that the average number of interactions within the continuum wormhole remains finite. Disconcertingly, as the P-CTC prescription lacks such a mechanism with which it may likewise be restrained, we determined that the median of its evolution probability distribution blows up as the number of modes diverges.

Considerably problematic, we discussed how the divergence of the clock expectation value in P-CTCs could stem from either a flaw in our model or a failing of the postselected teleportation prescription itself. Given that D-CTCs work well in the continuum limit, we are inclined to suspect the latter, and proposed a couple of methods by which one may suppress higher-order evolutions in the P-CTC solution. The first involved rendering the teleportation itself to be partial, or incomplete. Accomplished via the introduction of an independent parameter $\Incomplete$ into the entangled teleportation (Bell) state, it was shown that this method was sufficient in guaranteeing well-behaved probabilities in the continuum limit. Alternatively, the second method for ensuring convergence involved modifying the particle's self-interactions to be fundamentally probabilistic, such that the likelihood of a scattering event occurring was appointed a fixed value. Perhaps the most interesting facet of this result is that it suggests that in order to provide convergence in the continuum limit, P-CTCs must scale the collision probability inversely with the number of modes $\NumberModes$. This is in striking contrast with D-CTCs, for which we showed that the vacuum probability must scale directly with $\NumberModes$.

Inspired by previous work on CTCs and their reconciliation with quantum mechanics, the model established in this paper serves as an expeditionary foray into quantum time-travel paradoxes. The emergence of quantum solutions indicates that, like its classical counterpart, the quantum billiard-ball paradox is not innately pathological, nor is it ill-posed. Despite its simplicity, our model could serve as a useful basis for future work into similar problems, both classical and quantum.

\begin{acknowledgments}
	This research was supported by the Australian Research Council (ARC) under the Centre of Excellence for Quantum Computation and Communication Technology (Project No.~CE170100012). F.C.~acknowledges support through an ARC Discovery Early Career Researcher Award (DE170100712) and under the Centre of Excellence for Engineered Quantum Systems (EQUS, CE170100009). The University of Queensland (UQ) acknowledges the Traditional Owners and their custodianship of the lands on which UQ operates.
\end{acknowledgments}

\bibliographystyle{apsrev4-2}
\interlinepenalty=10000
\bibliography{billiard-ball_paradox_for_a_quantum_wave_packet}
	
\end{document}